\documentclass[sn-mathphys-num]{sn-jnl}

\usepackage{graphicx}%
\usepackage{multirow}%
\usepackage{amsmath,amssymb,amsfonts}%
\usepackage{amsthm}%
\usepackage{mathrsfs}%
\usepackage[title]{appendix}%
\usepackage{xcolor}%
\usepackage{textcomp}%
\usepackage{manyfoot}%
\usepackage{booktabs}%
\setcitestyle{numbers}
\usepackage{algorithm2e}
\usepackage{algorithmicx}%
\usepackage{algpseudocode}%
\usepackage{listings}%
\usepackage{array}
\usepackage{mathrsfs}
\usepackage{mdwmath}
\usepackage{changepage}
\usepackage{cuted}
\usepackage{mathbbol}
\usepackage{mdwtab}
\usepackage{eqparbox}
\usepackage{algpseudocode,float}
\usepackage{threeparttable}
\usepackage{lipsum}
\usepackage{url}
\usepackage{subfigure}
\usepackage{verbatim}
\usepackage{natbib}
\usepackage{booktabs}

\usepackage{caption}
\usepackage{makecell}
\usepackage{graphicx}
\usepackage{textcomp}
\usepackage{xcolor}
\usepackage{url}
\usepackage{amsmath,amssymb,amsfonts}
\usepackage{array}
\usepackage{color,framed}
\usepackage{multirow}
\usepackage{ulem}  
\usepackage{setspace}
\usepackage{comment}
\usepackage{pifont}

\def\BibTeX{{\rm B\kern-.05em{\sc i\kern-.025em b}\kern-.08em
    T\kern-.1667em\lower.7ex\hbox{E}\kern-.125emX}}
\usepackage{amsmath}
\usepackage{graphicx}
\usepackage{pgfplots}

\theoremstyle{thmstyleone}%

\theoremstyle{thmstyletwo}%

\theoremstyle{thmstylethree}%

\begin{document}
\title[Detecting Coverage Holes in Wireless Sensor Networks Using Connected Component Labeling and Force-Directed Algorithms]{\textcolor{black}{Detecting Coverage Holes in Wireless Sensor Networks Using Connected Component Labeling and Force-Directed Algorithms}}


\author[1]{ \sur{Jiacheng Xu}}\email{mc05426@um.edu.mo}

\author[1]{ \sur{Xiongfei Zhao}}\email{yb97480@um.edu.mo}

\author[2]{ \sur{Hou-Wan Long}}\email{houwanlong@link.cuhk.edu.hk}

\author[3]{ \sur{Cheong Se-Hang}}\email{dit.dhc@lostcity-studio.com}

\author*[4]{ \sur{Yain-Whar Si}}\email{fstasp@um.edu.mo}

\affil[1, 3, 4]{\orgdiv{University of Macau, Macau}}

\affil[2]{\orgdiv{The Chinese University of Hong Kong, Hong Kong}}

\abstract{Contour detection in Wireless Sensor Networks (WSNs) is crucial for tasks like energy saving and network optimization, especially in security and surveillance applications. Coverage holes, where data transmission is not achievable, are a significant issue caused by factors such as energy depletion and physical damage. Traditional methods for detecting these holes often suffer from inaccuracy, low processing speed, and high energy consumption, relying heavily on physical information like node coordinates and sensing range. To address these challenges, we propose a novel, coordinate-free coverage hole detection method using Connected Component Labeling (CCL) and Force-Directed (FD) algorithms, termed FD-CCL. This method does not require node coordinates or sensing range information. We also investigate Suzuki's Contour Tracing (CT) algorithm and compare its performance with CCL on various FD graphs. Our experiments demonstrate the effectiveness of FD-CCL in terms of processing time and accuracy. Simulation results confirm the superiority of FD-CCL in detecting and locating coverage holes in WSNs.}


\maketitle

\section{Introduction}
\label{section1}
A wireless sensor network (WSN) consists of static sensor nodes or mobile wireless terminals placed within a designated area. Its purpose is to continually transmit data in applications such as environment monitoring, intrusion detection, and surveillance systems~\cite{akyildiz2002wireless}. These sensor nodes are typically deployed randomly or manually in the case of surveillance applications, to collect, process, and transmit various types of sensed data such as movement path, signal strength, and number of connections. There are several research problems in WSN applications, including coverage, heterogeneity, security, synchronization, and data transmission~\cite{yick2008wireless}. In this paper, we focus on the coverage hole detection problem in WSN research. 

Hole detection is a crucial research topic in the field of wireless sensor computing, with wide-ranging applications in emergency response, battlefield rescue, and surveillance. In these contexts, accuracy and processing speed are essential factors for the effective detection of sudden events~\cite{zhao2016detecting}. Moreover, Hole detection in WSN graphs have been shown to have significant benefits, including reducing energy consumption for tracking mobile targets, optimizing network structure, and improving readability through the properties of contour. 

Existing methods for detecting contours in WSN can be primarily classified into three groups: geometry-based, topology-based, and statistical approaches. Geometry-based methods require accurate location information of sensors, such as GPS, and use computational geometry techniques such as Voronoi Diagram and Delaunay Triangulation to trace contours in the network. While this approach can precisely detect all coverage holes in the region of interest (ROI), obtaining accurate location information is limited by network delays and high energy consumption~\cite{yan2020connectivity}. Topology-based approaches only require connectivity information and distance between connected sensor nodes, making them easier to obtain. Homology theory-based algorithms have received the most attention in this group, as they study topological characteristics of networks using algebraic tools to calculate the number and location of coverage holes. In the third group, sensor nodes are deployed uniformly in the ROI, and detection methods rely on statistical properties such as node distribution probability, energy level for enhanced transmission, and specific network models~\cite{das2020review}. Furthermore, computer vision with force-directed(FD) based hole detection method uses FD graph to simulate WSN and applies contour tracing method~\cite{suzuki1985topological} on FD graph to detect holes. With the recent advancement in machine learning, hole detection methods based on convolution neural network(CCN) and FD were proposed in~\cite{2022An},~\cite{0Coverage}, which use graphs with detected holes to train hole detection model.

However, these contour tracing methods constantly encounter three fundamental problems from hole detection in practical WSN graphs. First, since due to the dynamic nature of WSN graphs, it has a high probability that the result after processing will be inaccurate. 
Second, decreasing the connectivity of WSN or unbalancing the load of a WSN can effect the performance~\cite{funke2006hole}. According to the properties of WSN graph, such as non-uniformity of deployment~\cite{vales2013optimal}, unstable signal of sensor nodes~\cite{zhu2012survey} and obstacle of terrain~\cite{benkirane2012performance}, interference points are bound to appear and progressively increase in number and size~\cite{zhao2016detecting}. Last but not the least, calculating the area of contour and producing the mapped contour graph from dense network only by sensor node's properties is impossible because the network has overlapping lines produced by sensor nodes. It is possible to get erroneous results when an interfering moving sensor node appears on the contour~\cite{koriem2020detecting}.
%
To alleviate these problems, in this paper, we propose a novel hole detection method with connected component labeling on FD graph. The proposed method can detect holes without using location information and the radius of circular sensing range of sensor nodes. 
The contributions of this paper are summarized as follows:
\begin{itemize}
	\item We propose a novel hole detection method called FD-CCL for detecting holes in WSNs.
	\item FD-CCL can detect holes without using information such as physical locations of the nodes and the radius of circular sensing range of sensor nodes.
	\item Extensive experiments were conducted to evaluate the performance of FD-CCL.  Experimental results on comparison with latest FD-based hole detection algorithm (FD-CT) confirm the superiority of FD-CCL in detecting and locating coverage holes in WSNs.
\end{itemize}

This paper is organised as follows: Section~\ref{section2} reviews related work. In Section~\ref{section3}, we briefly introduce the algorithms used in the proposed approach. Section~\ref{section4} describes our proposed method (FD-CCL) for hole detection in WSN. Experimental results are discussed in Section ~\ref{section5}. We conclude the paper with future work in Section ~\ref{section6}.

\section{Related Work}
\label{section2}

\subsection{Hole Detection in WSN}
Hole detection with contour tracing is one of the challenging research topics  due to the unpredictable network structure of WSN and overlapping lines. Furthermore, contour recognition is one of the most basic pre-processing operations in a vast range of WSN-based applications, from various monitoring systems to optimize energy consumption. Hole detection based on contour tracing is performed when the detection program needs to recognize contour pixels of the holes from 2D wireless sensors network image. In this section, we review related work on localization algorithms as well as on hole-detection. Hole detection methods can be classified into several categories \cite{das2020review}.

Topological based hole detection methods utilize connectivity information to identify contour nodes without the need for additional location identification devices, such as GPS, resulting in minimized costs~\cite{4531889},~\cite{5449912}. However, collecting neighboring node information may result in high control packet overhead. Despite this, topological methods have moderate accuracy in detecting contour nodes compared to statistical methods. Banafsj Khalifa et al. developed a distributed algorithm called DHDR, which aims to repair coverage holes in a sensor network using only the nodes already present in the network. The algorithm employs mobile nodes and does not require outside coordination~\cite{khalifa2021distributed}. Moreover, in~\cite{yan2020connectivity}, connectivity-based k-coverage hole detection algorithm was proposed by FengYan et al. This approach uses the Rips complex in homology theory to model a wireless sensor network, enabling accurate detection of over 95\% of non-triangular coverage holes.

Computational Geometry (CG) based hole detection method uses positioning technology, such as GPS or cost-effective geometry-based localization methods to collect node location information. The CG approach is highly accurate in detecting holes compared to topological and statistical approaches~\cite{kanno2009detecting},~\cite{kang2013detection}. A modified method for hole detection and healing was proposed by Kumar et al. The proposed method uses a distributed virtual force-based approach to remove redundant nodes and relocate nodes to eliminate holes. This energy and cost-efficient method provides a virtual force-based strategy for efficient hole detection and healing~\cite{kumar2015coverage}.

Methods based on statistical properties have less accuracy in comparison to topology and CG based approaches,~\cite{senouci2013localized},~\cite{kumar2016efficient}. In this case, sensor nodes are assumed to be distributed uniformly in the region of interest (ROI) and location information of the nodes are unavailable. Therefore, contour nodes are identified based on statistical property. This approach has less accuracy in comparison to topology and CG based approaches~\cite{aliouane2016efficient}. Yunzhou Zhang et al. introduced a new coverage hole detection method called HDRE for randomly deployed wireless sensor networks. This method incorporates residual energy to prevent coverage holes from reappearing soon after repairing the network~\cite{zhang2014hdre}.

A hole detection method called FD-CNN which is based on force-directed algorithms with convolution neural network (CNN) was proposed in~\cite{2022An}. In this method, force-directed algorithms are used to generate the layout of WSNs. Next, a CNN model was trained using the training dataset containing the rectangle area and the centroid of the labeled holes~\cite{2022An}. The model was then used to detect coverage hole from the layout. In~\cite{2022Hole}, Se-Hang Cheong et al. has developed an approach called FD-CT for detecting holes in wireless sensor networks by exploiting the power of force-directed algorithms for generating layouts from a given topology. The generated layouts are then processed with a contour tracing algorithm for identifying the sensors along the pixels of holes in the networks~\cite{2022Hole}. To address the problem of insufficient training data, a hole detection method named FD-TL was proposed in~\cite{0Coverage}. FD-TL was based on force-directed algorithms and transfer learning. The proposed approach applied Force-Directed algorithms for generating the layout of WSNs and CNN model for detecting both triangular and non-triangular holes without relying on the physical information of sensors. In their method, transfer learning was used to solve the problem of insufficient training data~\cite{0Coverage}.

In this paper, we propose a novel hole detection method which is based on  connected component labeling (CCL) based contour tracing algorithm. CCL was also used in image pre-processing procedure of computer vision applications. Different from the above hole detection methods, CCL based contour tracing method can extract contour pixels without relying on machine learning model, which can be a practical solution to deal with realistic WSN graphs. It is worth mentioning that fast and simple image processing algorithms can be beneficial in hole detection due to the limited hardware resources of mobile sensor devices. By exploiting the advantages of CCL algorithm such as high processing velocity and low memory usage~\cite{seo2016fast}, we can obtain the properties of contour such as height, width, area and coordinates in an highly efficient way.

\section{Background}
\label{section3}

\subsection{Force-Directed Graph Visualization Algorithm}

\begin{figure}[htbp]
	\centering
	\includegraphics[width=0.8\textwidth]{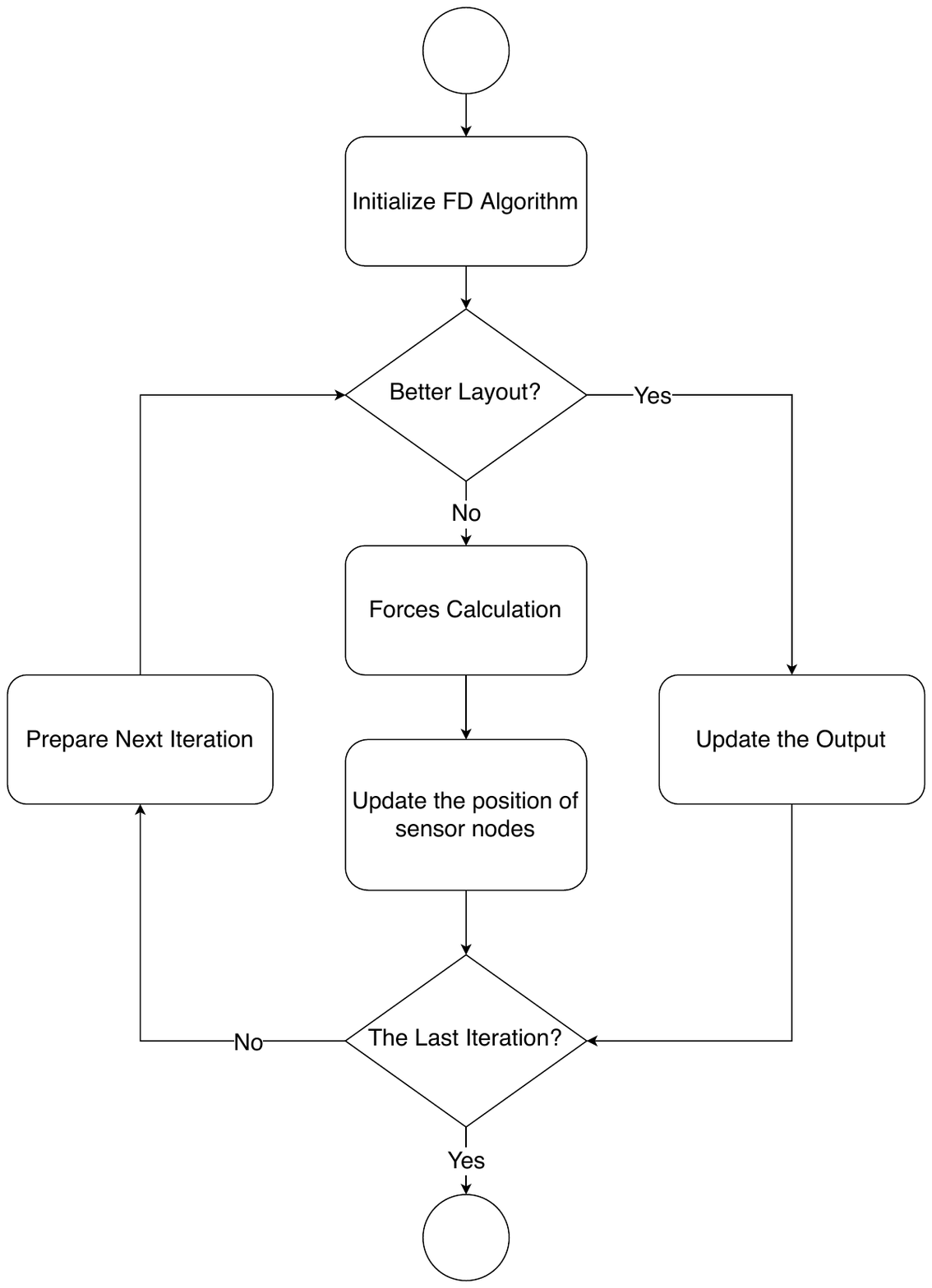}
	\caption{The control flow of a typical FD algorithm.}
	\label{fig:FD_ControlFlow}
\end{figure}

Force-Directed (FD) algorithm is a kind of graph visualization technique developed in 1960s. FDs are widely used in various applications such as information visualization, routing optimization, sensor networks, and graph drawing~\cite{cheong2020force}. The FD algorithm initially places nodes in random positions and iteratively moves them until the forces among them are in balance. The forces are calculated based on rules that prevent nodes from overlapping and encourage edge connections~\cite{komarek2015network}. The algorithm stops iterating when the layout properties satisfy the criteria, resulting in a visually appealing representation. The main control flow of a typical FD algorithm is illustrated in Figure~\ref{fig:FD_ControlFlow}.

In this paper, we focus on the FD algorithms to produce readable and aesthetically pleasing graphs of WSNs without requiring sensor node locations while minimizing edge-crossings. In the experiments, we generated uniform and sparse layouts from a list of edges using various FD algorithms, such as the Kamada-Kawai (KK) algorithm~\cite{kamada1989algorithm}, JIGGLE algorithm~\cite{tunkelang1998jiggle}, ForceAtlas2 (FA2) algorithm~\cite{jacomy2014forceatlas2}, and Fruchterman Reingold (FR) algorithm~\cite{fruchterman1991graph}.

\subsubsection{Kamada-Kawai Force-Directed Algorithm}
The Kamada-Kawai algorithm is a force-directed graph layout algorithm that aims to minimize the total energy of a graph by treating it as a physical system~\cite{kamada1989algorithm}. The algorithm calculates the forces between nodes and iteratively moves them until the system's total energy is minimized. The force calculation equations between node $i$ and $j$ are defined as follows:
\begin{itemize}
	\item [1)]
	Initial edge length:
	\begin{equation}
		L(i, j) = k * (d(i, j) - r(i) - r(j))
	\end{equation}
	
	\item [2)]
	Total energy of system:
	\begin{equation}
		E = sum((L(i, j) - d(i, j))^2 / L(i, j))
	\end{equation}
	\item [3)]
	Force between node $i$ and node $j$:
	\begin{equation}
		F(i, j) = k * (d(i, j) - L(i, j)) * (x(i) - x(j)) / d(i, j)
	\end{equation}
	\item [4)]
	Position update:
	\begin{equation}
		pos(i) = pos(i) + delta_t * sum(F(i, j))
	\end{equation}
\end{itemize}
where $k$ is a constant that determines the overall stiffness of the edges, $d(i,j)$ is the distance between node $i$ and node $j$, and $r(i)$ represents the radius of node $i$.

In each iteration, the algorithm moves the nodes to reduce the cost function until it reaches a minimum. The algorithm stops when the cost function can no longer be reduced or when a maximum number of iterations is reached. The KK algorithm produces layouts that are aesthetically pleasing and easy to interpret, with short edges and non-overlapping nodes. However, it can be computationally expensive for large graphs and may not always produce the best layout for certain types of graphs.

\subsubsection{ForceAtlas2 Force-Directed Algorithm}
ForceAtlas2 is a force-directed layout algorithm used to visualize graphs, introduced by Jacomy et al.~\cite{jacomy2014forceatlas2}. The algorithm uses a combination of attractive and repulsive forces to position the nodes in the graph, aiming to achieve a layout that minimizes the system's energy. The equations used by ForceAtlas2 to calculate the forces acting on each node are as follows:
\begin{itemize}
	\item [1)]
	Repulsive force:
	\begin{equation}
		F_{rep}(i,j) = K_{rep} * \frac{(deg(i)+1) * (deg(j)+1)}{d_{ij}}
	\end{equation}
	\item [2)]
	Attractive force:
	\begin{equation}
		F_{att}(i,j) = d_{ij}
	\end{equation}
	\item [3)]
	Gravity force:
	\begin{equation}
		F_{gravity}(i) = K_g * (deg(i) + 1)
	\end{equation}
\end{itemize}
where $K_{rep}$ and $K_g$ are scaling factors, $deg(i)$ is the degree of node $i$, and $d_{ij}$ is the distance between nodes $i$ and $j$. Additionally, ForceAtlas2 uses a damping factor to control the speed of layout convergence, and the gravity force $F_{gravity}$ prevents disconnected components from drifting away.

\subsubsection{Fruchterman Reingold Force-Directed Algorithm}
The Fruchterman-Reingold (FR) algorithm is a force-directed graph layout algorithm proposed by Fruchterman et al.~\cite{fruchterman1991graph} and is commonly used to visualize complex networks. The algorithm is based on the physical analogy of a system of particles connected by springs, where the goal is to minimize the system's energy by adjusting the particles' positions.

The FR algorithm models the graph layout as a physical system, with nodes representing particles and edges representing springs. It seeks to minimize the total energy, defined as the sum of the potential and kinetic energies of the particles. The algorithm applies two types of forces: attractive forces between adjacent nodes and repulsive forces between non-adjacent nodes. The strength of these forces is determined by a constant $k$ and the Euclidean distance between the nodes.

The total force acting on each node is the sum of the attractive and repulsive forces. The displacement of each node is calculated based on this total force, with a temperature parameter $T$ controlling the rate of movement in the simulation. The FR algorithm iteratively updates the positions of the nodes based on these forces until the system reaches equilibrium. The equations for the forces and the node displacement are as follows:
\begin{itemize}
	\item [1)]
	Repulsive force between nodes $i$ and $j$:
	\begin{equation}
		F_{rep}(i,j) = \frac{k_r}{d_{ij}^2}
	\end{equation}
	\item [2)]
	Attractive force between nodes $i$ and $j$:
	\begin{equation}
		F_{att}(i,j) = k_a \cdot \log_{}{d}
	\end{equation}
	\item [3)]
	Total force:
	\begin{equation}
		F(i) = \sum_{j \neq i}^n \left(F_{rep} - F_{att}\right)
	\end{equation}
	\item [4)]
	The displacement of node $i$ is given by:
	\begin{equation}
		\Delta x_i = \frac{F_i}{|F_i|} \cdot \min(F_i, T)
	\end{equation}
	\item [5)]
	The new position of node $i$ is given by:
	\begin{equation}
		x_i^{(t+1)} = x_i^{(t)} + \Delta x_i
	\end{equation}
\end{itemize}
where $k_r$ and $k_a$ are constants, $d_{ij}$ is the Euclidean distance between nodes $i$ and $j$, $n$ is the total number of nodes in the network, $T$ is the temperature parameter, and $x_i^{(t)}$ is the position of node $i$ at iteration $t$. By iteratively applying these equations until the system reaches equilibrium, the FR algorithm produces a visually appealing and informative layout of the graph.

\subsubsection{JIGGLE Force-Directed Algorithm}

The Java Interactive Graph Layout Environment (JIGGLE) proposed by Tunkelang et al.~\cite{tunkelang1998jiggle} is a force-directed algorithm specifically designed for large graphs. It employs several optimizations to enhance the layout process efficiency, such as reducing the number of pairwise interactions between nodes and using a multilevel approach to partition the graph into smaller subgraphs. The equations used by JIGGLE to compute the forces between nodes are as follows:
\begin{itemize}
	\item [1)]
	Repulsive force between nodes i and j:
	\begin{equation}
		F_{rep}(i,j) = k_r\frac{(x_i - x_j)}{\|x_i - x_j\|} \cdot \frac{1}{\|x_i - x_j\|}
	\end{equation}
	\item [2)]
	Attractive force between nodes i and j:
	\begin{equation}
		F_{att}(i,j) = k_a\frac{(x_i - x_j)}{\|x_i - x_j\|} \cdot \left(\log\frac{\|x_i - x_j\|}{l_{ij}}\right)
	\end{equation}
	\item [3)]
	Spring force along an edge e connecting nodes i and j:
	\begin{equation}
		F_{spring}(e) = k_s(l_{ij} - \|x_i - x_j\|)\frac{(x_i - x_j)}{\|x_i - x_j\|}
	\end{equation}
\end{itemize}
In these equations, $x_i$ and $x_j$ are the positions of nodes $i$ and $j$ in the layout, $l_{ij}$ is the desired length of the edge connecting nodes $i$ and $j$, and $k_r$, $k_a$, and $k_s$ are constants that control the strength of the repulsive, attractive, and spring forces, respectively.

\subsection{Connected Component Labeling}
Connected Component Labeling (CCL) is a pixel-following contour tracing algorithm that uses a chain code to represent image regions. CCL requires frame-sized memory for contour tracing and provides a compact region representation, widely used as a standard input format for numerous shape analysis algorithms. It traces contour pixels in a predefined manner and saves their coordinates in memory according to the trace order~\cite{he2017connected}. There are three steps in Connected Component Labeling Contour Tracing method: Threshold Processing, Connected Component Identification, and Labeling.

Threshold Processing is a pre-processing step that converts the input image into a binary image. Each pixel in the binary image is either 0 or 1, depending on whether the corresponding pixel in the input image is above or below a certain threshold value. In the equation, $T(x)$ represents the binary value of the pixel at location $x$ in the binary image, $I(x)$ is the intensity value of the pixel at location $x$ in the input image, and $\theta$ is the threshold value. If $I(x)$ is greater than $\theta$, then $T(x)$ is set to 1; otherwise, $T(x)$ is set to 0. This step can be considered label initialization. The pixel label calculation is shown as follows:
\begin{equation}
	T(x) = \begin{cases}
		1, & \text{if } I(x) > \theta \\
		0, & \text{otherwise}
	\end{cases}
	\label{equa:ThresholdProcessing}
\end{equation}

The next step is to identify the connected components in the binary image. A connected component is a set of foreground pixels that are connected to each other through their neighboring pixels. This is typically achieved using a connected component analysis algorithm, such as the Two-Pass algorithm \textcolor{black}{and the Union-Find data structure}. 

\textcolor{black}
{Union-Find is a tree data structure whose operations are used to keep track of the different labels assigned to connected components during the labelling process. The Find operation is used to determine the root label of a given component, while the Union operation is used to join two components with different labels. In the Find operation, starting from a given pixel $x$, we follow its parent pointers until we reach a pixel whose parent is itself, thus identifying the root of the component containing $x$. The formula of the Find operation is shown below:}
\begin{equation}
	\begin{aligned}
		\text{Find}(x) = \begin{cases}
			x,                      & \text{if } parent(x) = x \\
			\text{Find}(parent(x)), & \text{otherwise}
		\end{cases}
	\end{aligned}
	\label{equa:Find}
\end{equation}
\\textcolor{black}{After finding the parent pixel of all pixels, the union operation merges two components with root labels $x$ and $y$ and compares their ranks by estimating the height of the tree rooted at each pixel. The union operation is formulated as}
\begin{equation}
	\begin{aligned}
		\text{Rank}(x) = \text{The number of pixels in the set}
	\end{aligned}
	\label{equa:Union}
\end{equation}
\begin{equation}
	\begin{aligned}
		\text{Union}(x, y) = \begin{cases}
			x, & \text{if } rank(x) > rank(y)                                                               \\
			y, & \text{if } rank(y) > rank(x)                                                               \\
			x, & \text{otherwise, with } parent(x) \leftarrow y \text{ and } rank(y) \leftarrow rank(y) + 1
		\end{cases}
	\end{aligned}
	\label{equa:Union}
\end{equation}

The neighborhood analysis equation defines the neighborhood of a given pixel $x$ in the image~\cite{jahne1999handbook}. To label the current pixel, we can use 4-connectivity or 8-connectivity to scan the graph. In CCL, 4-connectivity considers the north and west neighbors of the current pixel, while 8-connectivity includes the north-east, north, north-west, and west neighbors.
%
The width of the image is denoted by $w$. The 4-connectivity and 8-connectivity neighborhood equations can be formulated as follows:
\begin{equation}
	\begin{aligned}
		N_4(x) = \{x - 1, x + 1, x - w, x + w\}
	\end{aligned}
	\label{equa:N4}
\end{equation}
\begin{equation}
	\begin{aligned}
		N_8(x) = \{(x-1, y-1), (x, y-1), (x+1, y-1), (x-1, y), \\
		(x+1, y), (x-1, y+1), (x, y+1), (x+1, y+1)\}
	\end{aligned}
	\label{equa:N8}
\end{equation}

Finally, each connected component is assigned a unique label, which serves as an identifier for that particular object in the image. This is typically achieved by assigning a different label to each connected component and updating the label of each pixel in the component with the corresponding label.
%
The steps of Two-Pass CCL (see Algorithm \ref{alg:CCL}) are as follows: 
\begin{itemize}
	\item [1)]
	Label Initialization: Initialize a label matrix to all zeros using Equation \ref{equa:ThresholdProcessing}, with the same dimensions as the binary image. This matrix will be used to assign unique labels to each connected component.
	\item [2)]
	First Pass: Starting from the top-left corner, iterate over all pixels in the binary image from left to right, top to bottom. For each pixel, check its value and the values of its neighbors to determine if it is part of a connected component. If it is, assign a temporary label to the connected component and update a label equivalency set. The rules for assigning new label to current pixel are shown below:
	\begin{itemize}
		\item Rule 1: If the labels of all neighbor pixels are 0, assign label to current pixel and label increment.
		\item Rule 2: If the non-zero values are the same in all four directions, then the current pixel label is the non-zero label of its neighbors.
		\item Rule 3: If there are two different labels for non-zero values in four directions, the current position label will be assigned as one of them and the two different labels will be recorded to equivalent set since they are concatenated.
	\end{itemize}
	\item [3)]
	Second Pass: Iterate over all pixels in the binary image again, and replace the temporary labels assigned in the first pass with their final label from the label equivalency table.
\end{itemize}

\begin{algorithm}[htb]
	\SetAlgoLined
	\LinesNumbered
	\KwIn{Binary image with pixel values}
	\KwOut{Labeled image with connected components assigned unique labels}
	\textbf{First pass:} Initialize empty equivalence table and label count; \\
	\For{each pixel $x$ in image}{
		\If{$x$ is foreground pixel}{
			\If{$N_4(x)$ is empty}{
				Assign new label to $x$;
				Add new entry to equivalence table with label as key and empty set as value;
				Increment label count;
			}
			\Else{
				Let $L$ be the smallest label in $N4(x)$;
				Assign $L$ to $x$;
				\For{each label in $N4(x)$, excluding $L$}{
					Add $x$ to the equivalence set for that label;
				}
			}
		}
	}
	\textbf{Second pass:} \\
	\For{each pixel $x$ in image}{
		\If{$x$ is foreground pixel}{
			Let $L$ be the label assigned to $x$ in the first pass;
			Let $S$ be the equivalence set for label $L$;
			\If{$S$ is not empty}{
				Let $L$ be the smallest label in $S$;
				Assign $L$ to $x$;
				\For{each label in $S$, excluding $L$}{
					Add $L$ to the equivalence set for that label;
				}
			}
		}
	}
	\caption{Two-pass connected component labeling algorithm}
	\label{alg:CCL}
\end{algorithm}

\begin{figure}[htbp]
	\centering
	\subfigure[Initialization: Threshold processing and assign labels for all pixels.]{
		\begin{minipage}[t]{0.45\linewidth}
			\centering
			\includegraphics[width=2in]{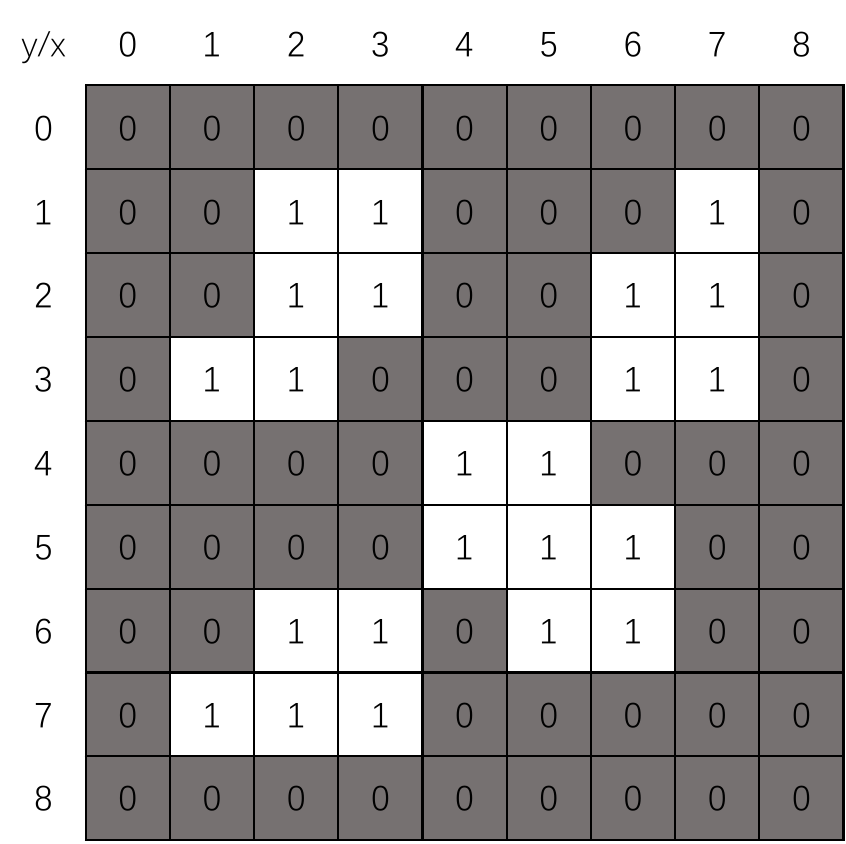}
		\end{minipage}
	}
	\subfigure[First pass phase: Scanning graph and update label.]{
		\begin{minipage}[t]{0.45\linewidth}
			\centering
			\includegraphics[width=2in]{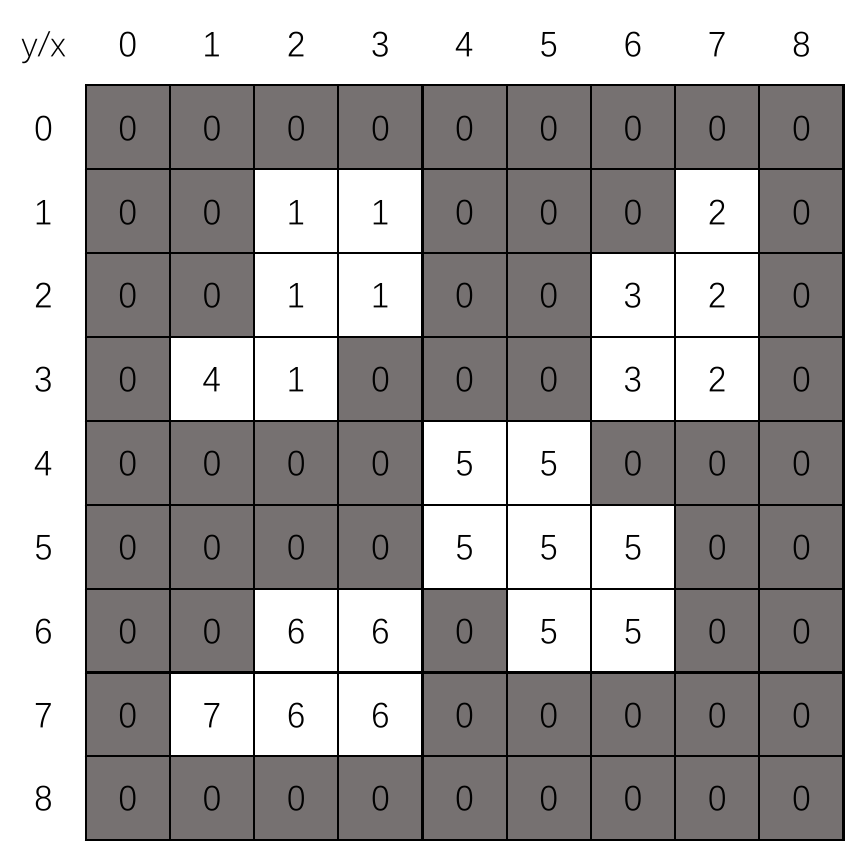}
		\end{minipage}
	}
	
	\subfigure[Equivalent set of each label.]{
		\begin{minipage}[t]{0.45\linewidth}
			\centering
			\includegraphics[width=2in]{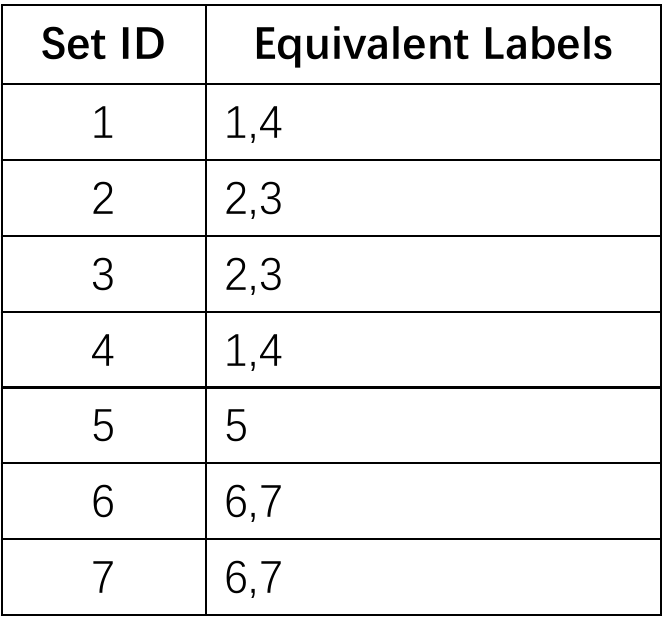}
		\end{minipage}
	}
	\subfigure[Second pass phase: Unified label based on equivalent set.]{
		\begin{minipage}[t]{0.45\linewidth}
			\centering
			\includegraphics[width=2in]{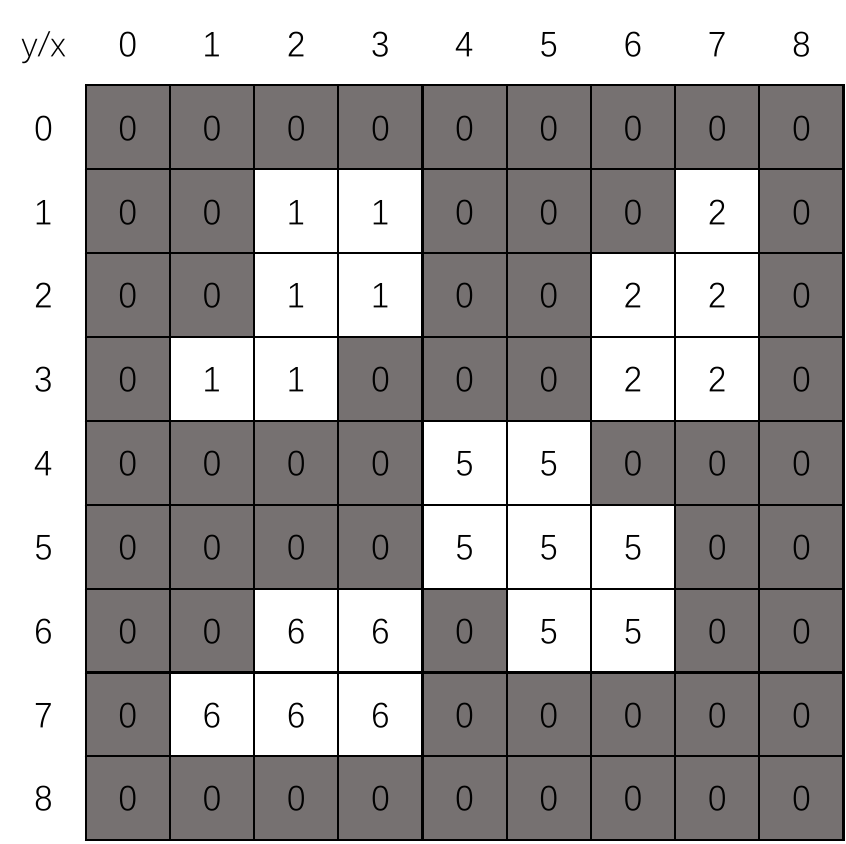}
		\end{minipage}
	}
	
	\subfigure[Color each area with identically label.]{
		\begin{minipage}[t]{0.45\linewidth}
			\centering
			\includegraphics[width=2in]{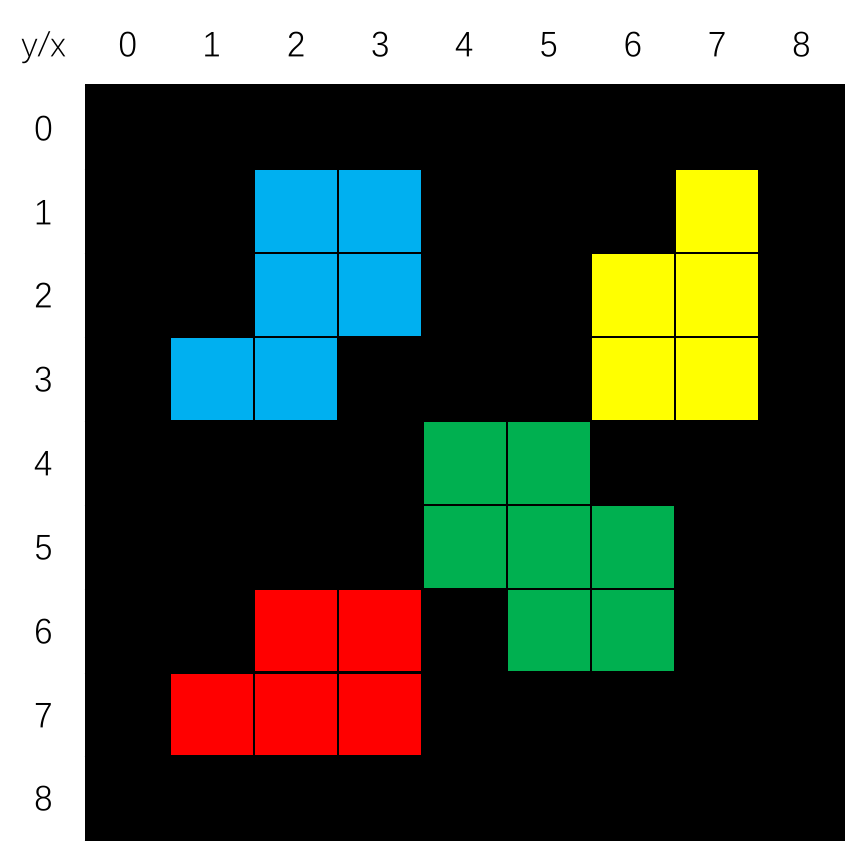}
		\end{minipage}
	}
	\centering
	\caption{Steps in Connected Component Labeling process.}
	\label{fig:CCL_MainProcedure}
\end{figure}

An example illustrating the steps of CCL is shown in Figure \ref{fig:CCL_MainProcedure}. $p_{xy}$ denotes the pixel at coordinates $(x, y)$, and $l_{xy}$ represents the label of $p_{xy}$. To initialize the graph, CCL treats black pixels as background pixels and assigns them a label of 0. White pixels are assigned with label 1, indicating they are foreground or contour pixels. In this example, we use a 4-connected neighborhood to scan the graph. In the first scan phase, suppose that $p_{xy} = (x_2, y_1)$ has no neighbors with a label greater than 0, then it is assigned an initial label of 1, and the label count is incremented. After the first pass phase, we obtain an equivalent set containing the same label list for each label. \\textcolor{black}{Based on the equivalent set (see Figure~\ref{fig:CCL_MainProcedure}(c)), we used the Union-Find algorithm to union the connected component. During the second pass phase, for example,} $l_{xy} = (x_1, y_3)$ \\textcolor{black}{will be assigned label 1} \cite{suzikiContourTracingExample}.

\section{Proposed Method (FD-CCL)}
\label{section4}
The proposed FD-CCL coverage hole detection method uses a combination of Connected Component Labeling (CCL) and Force-Directed (FD) graph visualization algorithms to identify holes in wireless sensor networks (WSN). As shown in Algorithm \ref{alg:FD}, with the input list of connectivity information from the source node to the target node, the layout of the WSN is generated by a FD algorithm. It is worth mentioning that layout generation step does not require any additional input information such as location of the nodes. Due to variations in FD calculations, different types of FD algorithms may generate different layouts of the WSN at a specific point in time given the input network topology.

\begin{algorithm}[htb]
	\SetAlgoLined
	\KwIn{Connectivity information of sensor nodes, Iteration times}
	\KwOut{Contours List, Properties List}
	\textbf{Initialization:} \\
	$C$ = connectivity information of sensor nodes; \\
	$S$ = the size of canvas; \\
	$t$ = minimum area threshold; \\
	$n$ = iteration times; \\
	$min_x,min_y$ = the minimum coordinates value in the $coordinates$; \\
	$O_x,O_y$ = the coordinates of sensor nodes in last iteration; \\
	$N_x,N_y$ = the coordinates of sensor nodes in current iteration; \\
	\For{\rm $1 \to n$}{
		\If{\rm the stable status threshold $r$ is less than $\varepsilon$}{
			continue; \\
		}
		$(N_x,N_y) = FD(O_x,O_y, C)$; \\
		$(N_x,N_y) = (O_x,O_y)+min(\frac{S}{min_x},\frac{S}{min_y})\times((N_x,N_y)-(O_x,O_y))$; \\
	}
	\Return{\rm $N_x,N_y$};
	\caption{Pseudo code of calling a FD algorithm}
	\label{alg:FD}
\end{algorithm}

\begin{figure}[htb]
 	\centering
 	\includegraphics[width=1.0\textwidth]{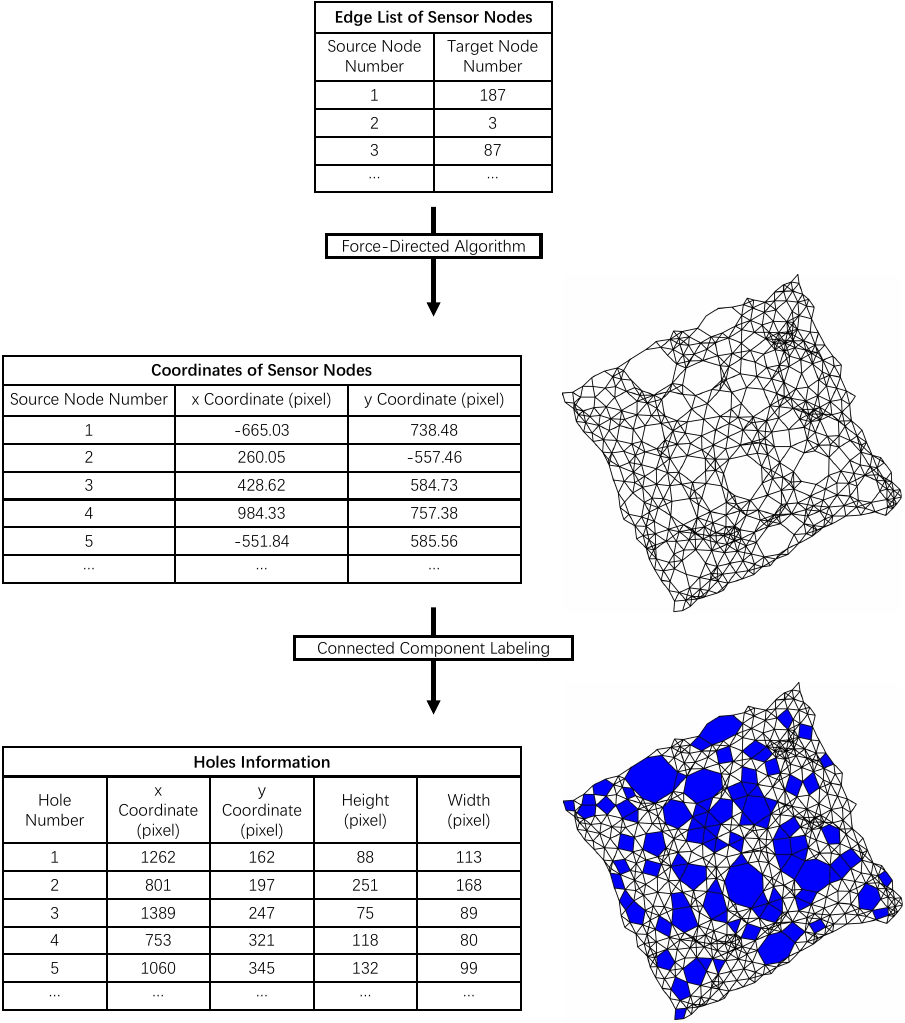}
 	\caption{The main procedure of FD-CCL.}
 	\label{fig:FD-CCL_MainProcedure}
 \end{figure}

 The overall procedure of FD-CCL is illustrated in Figure \ref{fig:FD-CCL_MainProcedure}. First, list of edges (connection information) are entered into a chosen Force-directed (FD) algorithm. Note that any FD algorithm reviewed in Section 3 can be chosen. Next, the FD algorithm will generate a layout (graph) containing the $x$ and $y$ coordinates of the sensor nodes. Note that these coordinates do not represent the actual physical coordinates of the sensor nodes. Rather, these coordinates are obtained from the final positions of the nodes from the layout generated by the FD algorithm using force calculation. After that, the generated layouts are exported into an image file or text format for post-processing. Finally, Connected Component Labeling (CCL) algorithm is used to detect the holes from the graph. CCL stores the coordinates, height, width, and area of the detected contours. Algorithm~\ref{alg:FD-CCL} summarizes the steps required for FD-CCL for detecting holes.

\begin{algorithm}[htb]
	\SetAlgoLined
	\KwIn{Connectivity information of sensor nodes}
	\KwOut{Contours List, Properties List}
	\textbf{Initialization:} \\
	$graph$ = result diagram; \\
	$t$ = minimum area threshold; \\
	$I_x,I_y$ = the initialization coordinates of sensor nodes; \\
	$R_x,R_y$ = the result coordinates of sensor nodes; \\
	$contours$ = empty list; \\
	$properties$ = empty list; \\
	$R_x,R_y$ = \textbf{FD}($I_x,I_y$); \\
	\textbf{Connected Component Labeling procedure:} \\
	$graph$ = $Pictorialization(R_x,R_y)$; \\
	$contours$, $properties$ = \textbf{CCL}($graph$); \\
	\ForEach{\rm $contour$ in $contours$}{
		\If{\rm the area of $properties(contour)$ < $t$}{
			$contours$ removes $contour$; \\
		}
	}
	\Return{\rm $contours$, $properties$};
	\caption{Pseudo code of FD-CCL algorithm}
	\label{alg:FD-CCL}
\end{algorithm}

\section{Experiments}
\label{section5}
In this section, we first evaluate the feasibility of the Force-Directed Connected Component Labeling Algorithm (FD-CCL) by detecting holes in a ground truth WSN with randomly deployed sensors. Secondly, we compare the performance of our FD-CCL approach and the Force-Directed Contour Tracing Algorithm (FD-CT) proposed by Cheong et al.~\cite{2022Hole}.

\subsection{Experiment Setup}
\begin{figure}[htp]
	\centering
	\subfigure[Nodes.]{
		\begin{minipage}[t]{0.3\linewidth}
			\centering
			\includegraphics[width=1.5in]{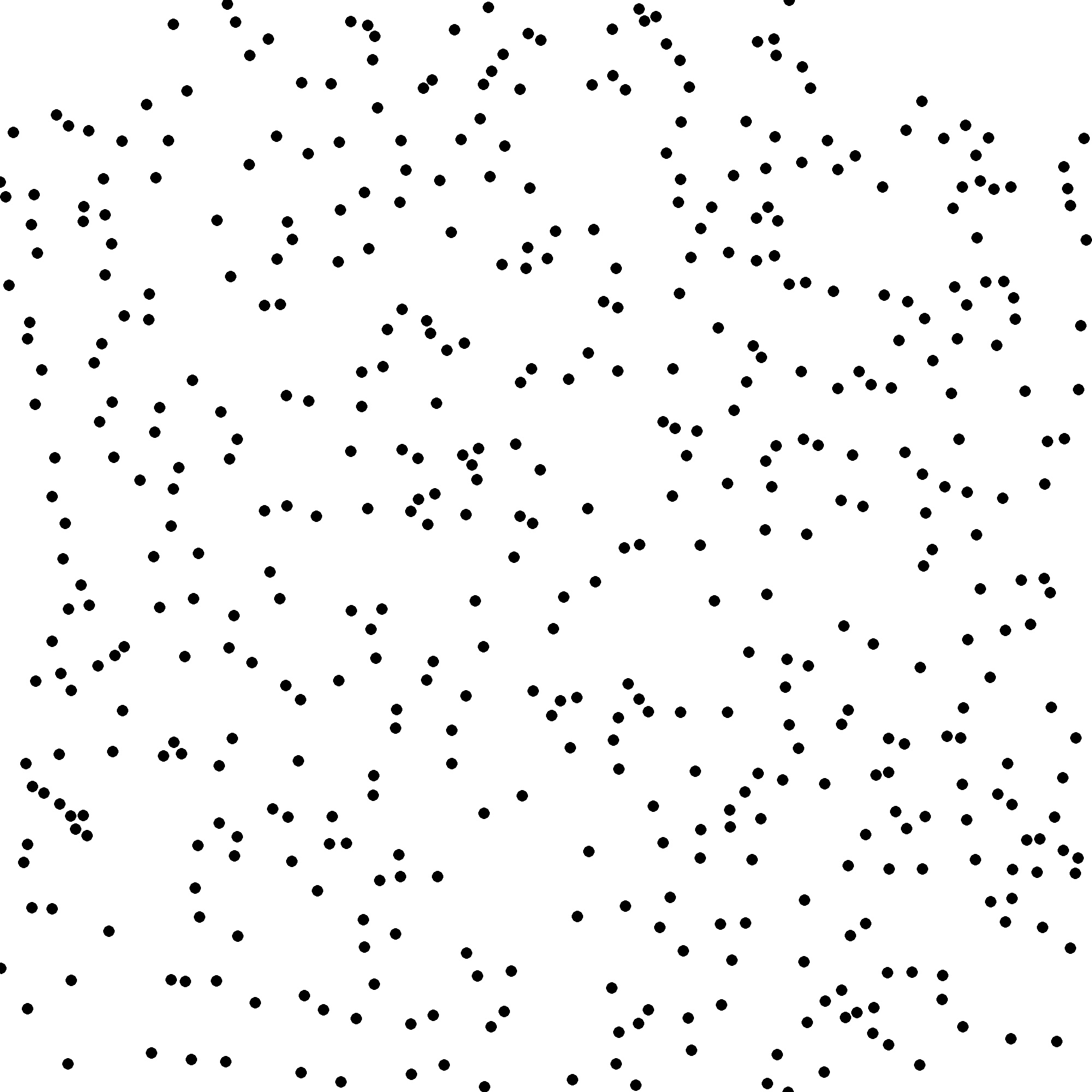}
		\end{minipage}
	}
	\subfigure[Edges.]{
		\begin{minipage}[t]{0.3\linewidth}
			\centering
			\includegraphics[width=1.5in]{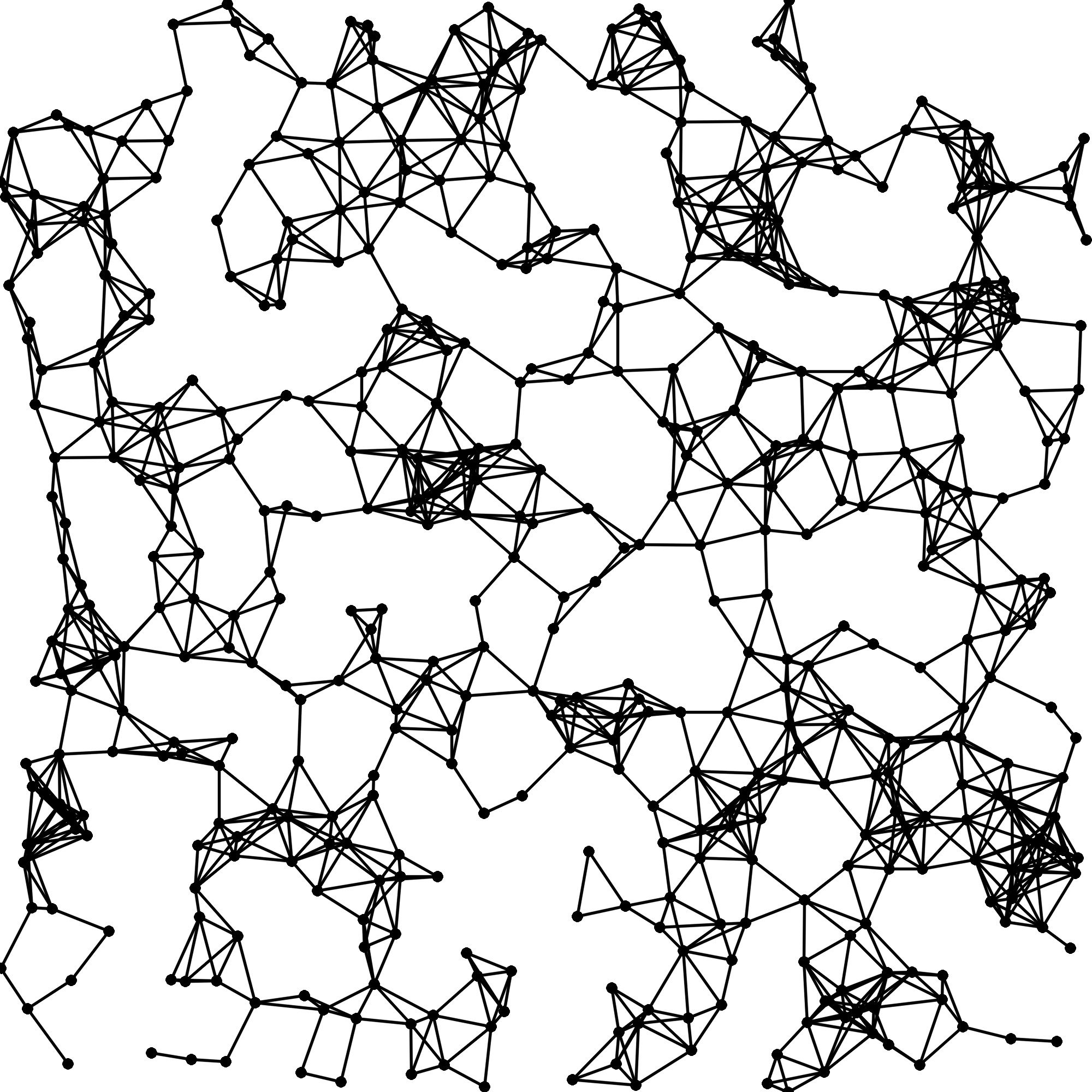}
		\end{minipage}
	}
	\subfigure[Contour tracing.]{
		\begin{minipage}[t]{0.3\linewidth}
			\centering
			\includegraphics[width=1.5in]{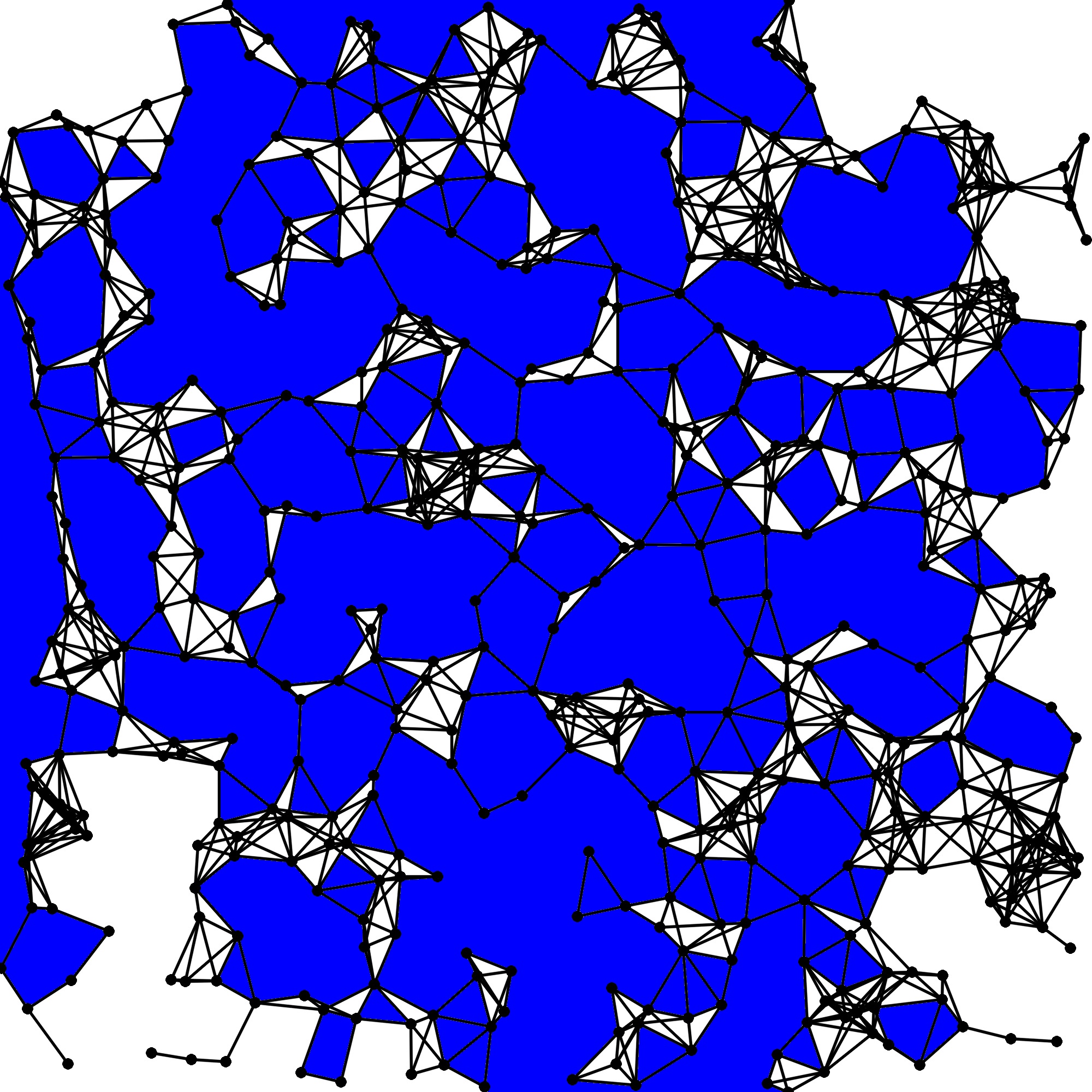}
		\end{minipage}
	}
	\caption{Examples of Ground Truth Dataset.}
	\label{fig:GroundTruth_Example}
\end{figure}

The examples of the ground truth dataset are shown in Figure \ref{fig:GroundTruth_Example}. Based on this ground truth dataset, we evaluate the performance of FD-CCL algorithm in Section \ref{subsection:FD-CCL_feasibility}.

\begin{figure}[htp]
	\centering
	\subfigure[Sparse layout.]{
		\begin{minipage}[t]{0.45\linewidth}
			\centering
			\includegraphics[width=2in]{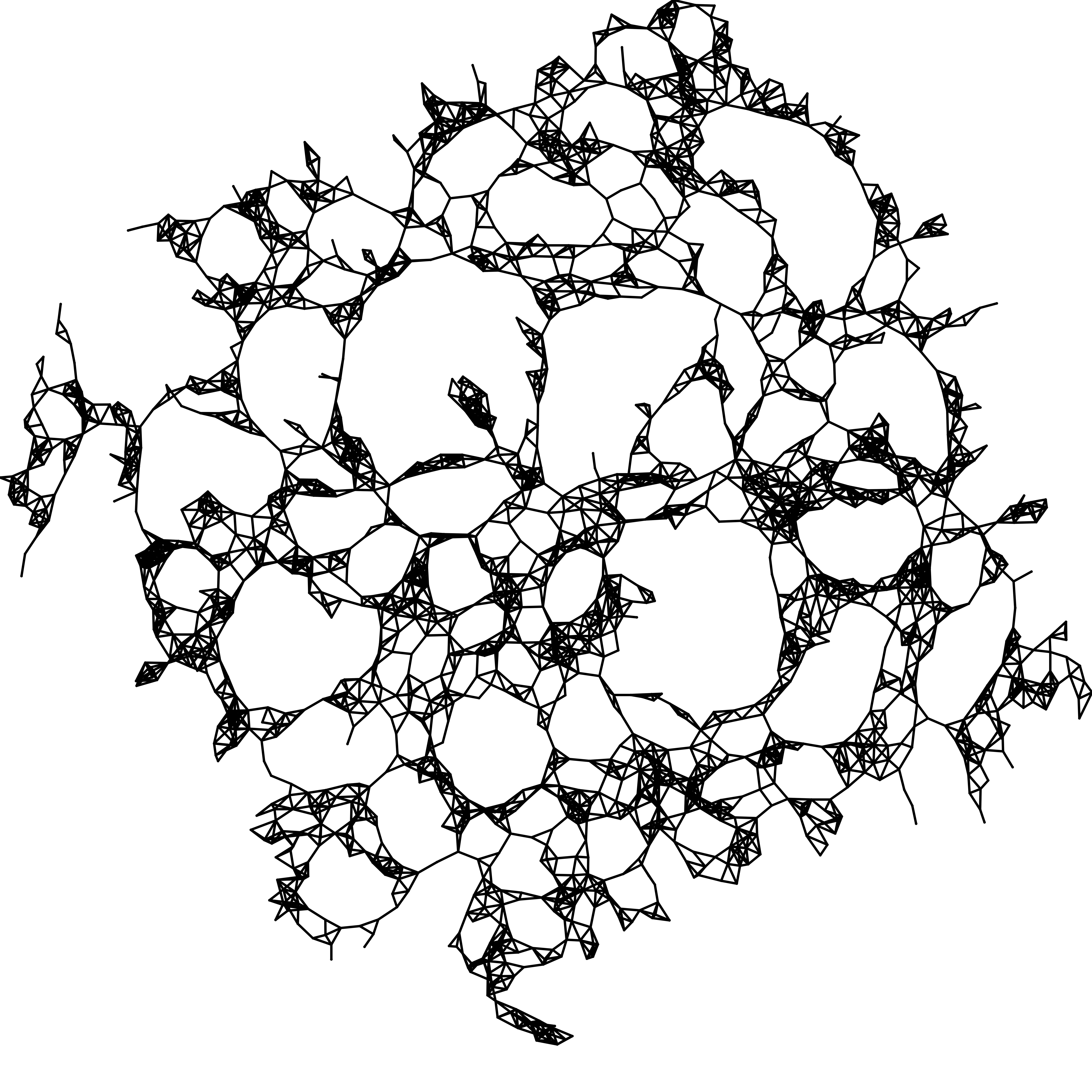}
		\end{minipage}
	}
	\subfigure[Uniform layout.]{
		\begin{minipage}[t]{0.45\linewidth}
			\centering
			\includegraphics[width=2in]{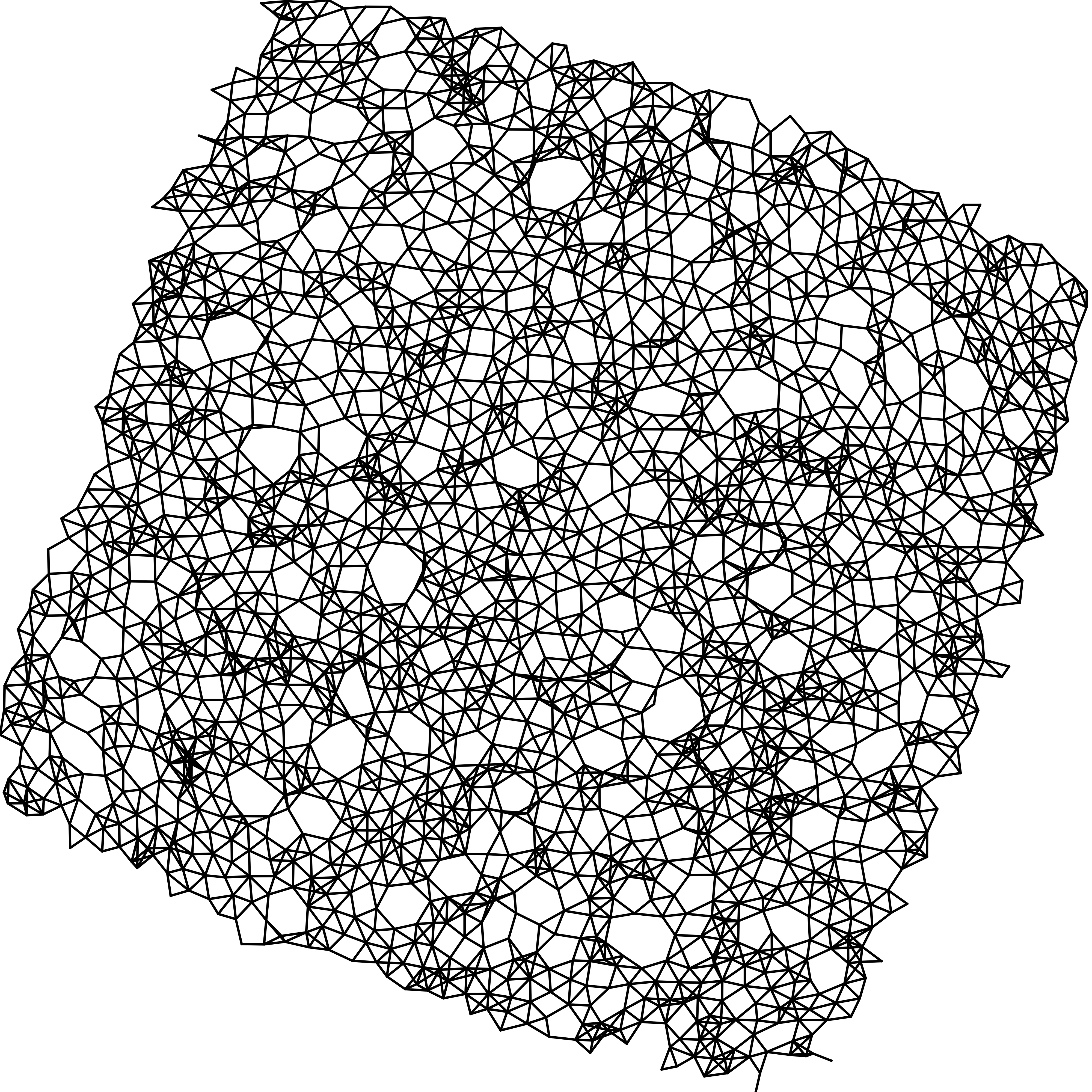}
		\end{minipage}
	}
	\caption{Examples of sparse and uniform layouts generated by force-directed algorithms.}
	\label{fig:Sparse_Uniform}
\end{figure}

To evaluate the effectiveness of the FD-CT and FD-CCL approaches, we developed a simulation environment in Python. 
We conducted simulations with different numbers of nodes $n$ (200, 500, 1000, 2000, and 3000) and set the average degrees of sensors $d$ to 6 and 15, resulting in sparse and uniform layouts from a list of edges. We evaluated hole detection in a total of four WSN topologies based on all possible combinations of $n$ and $d$. Figure \ref{fig:Sparse_Uniform} illustrates the examples of sparse and uniform layouts generated by FD algorithms.

\begin{figure}[htp]
	\centering
	\includegraphics[width=0.8\textwidth]{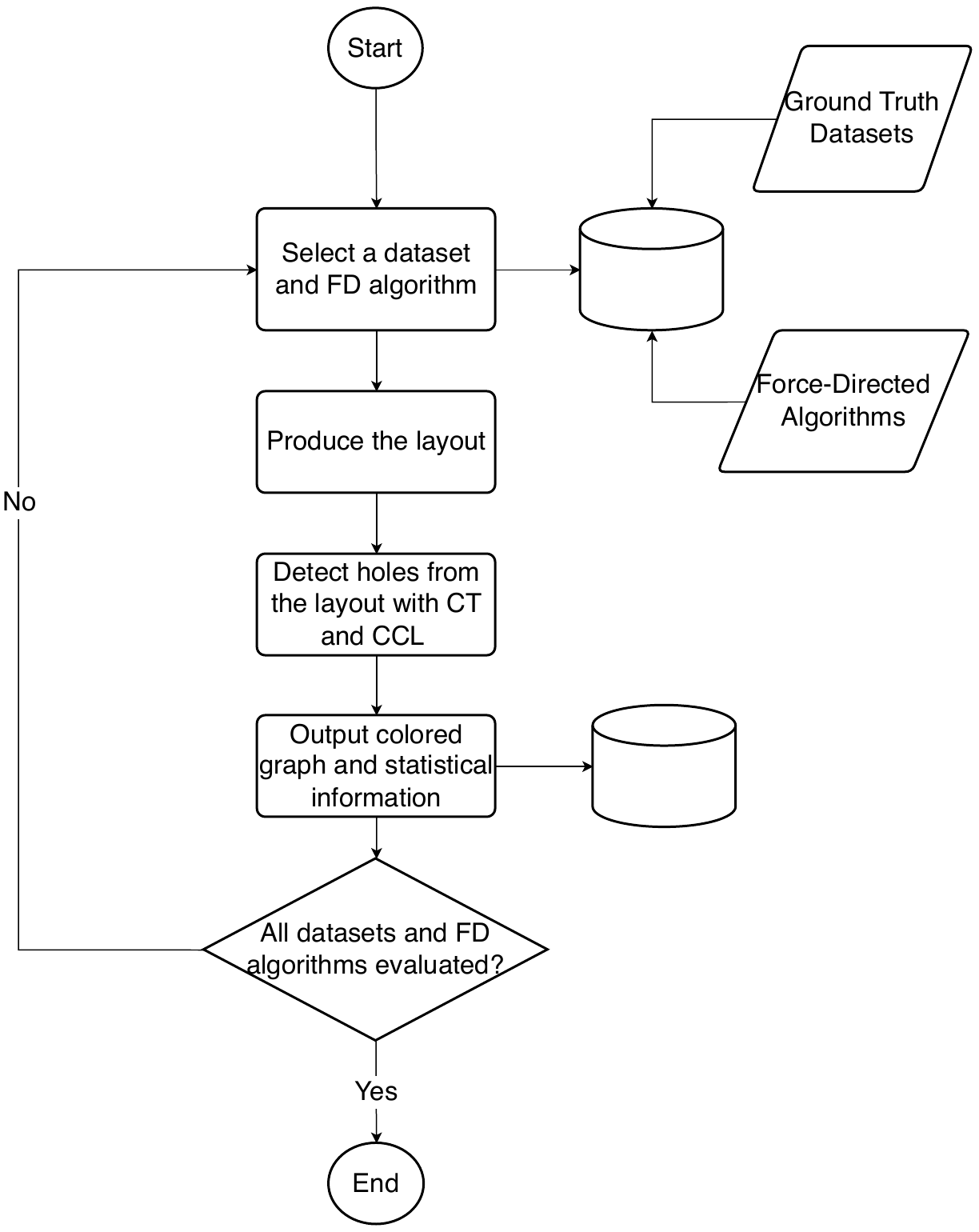}
	\caption{The experiment workflow of FD-CCL and FD-CT.}
	\label{fig:Experiment_Workflow}
\end{figure}

\begin{figure}[htp]
	\centering
	\subfigure[FD-CT.]{
		\begin{minipage}[t]{0.45\linewidth}
			\centering
			\includegraphics[width=2in]{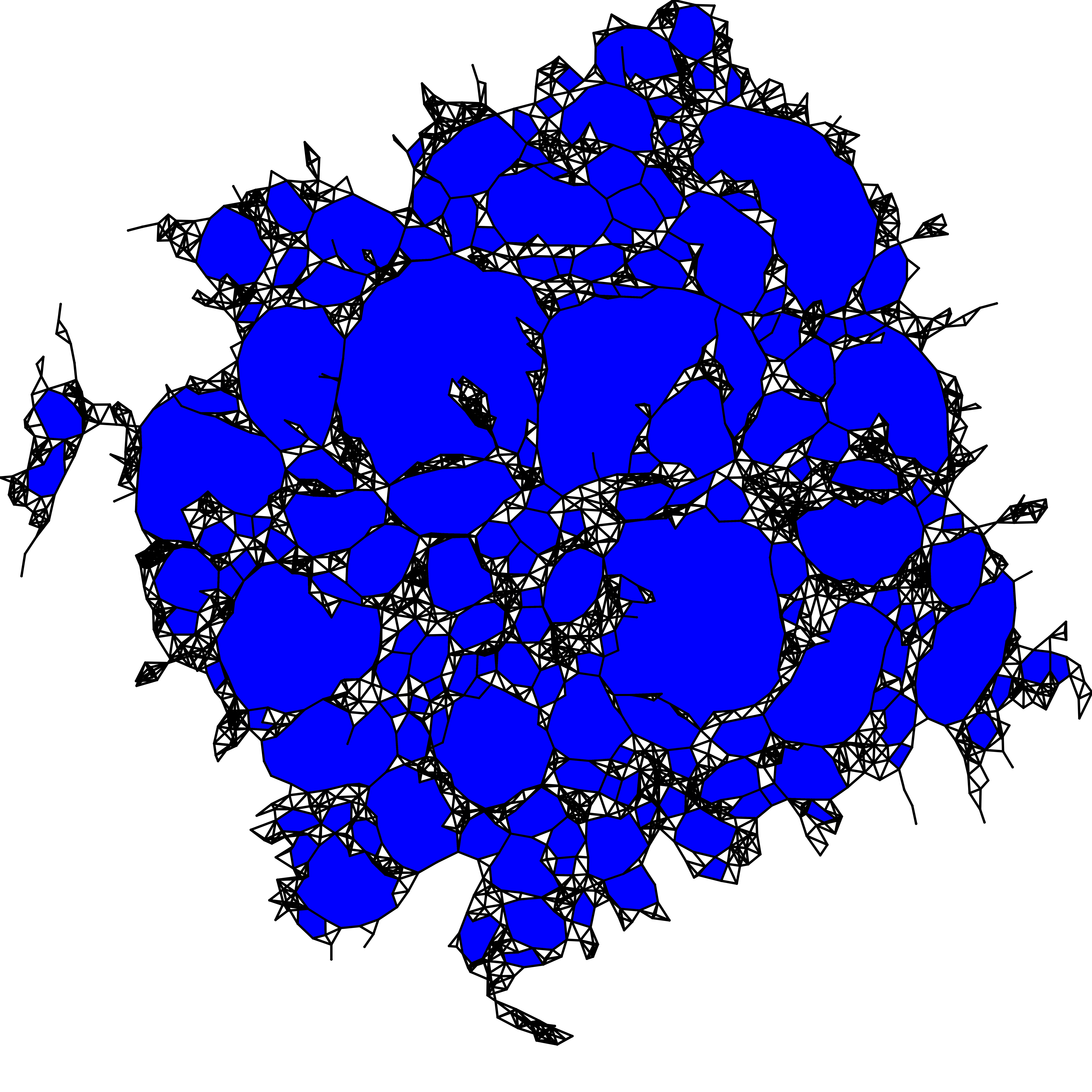}
		\end{minipage}
	}
	\subfigure[FD-CCL.]{
		\begin{minipage}[t]{0.45\linewidth}
			\centering
			\includegraphics[width=2in]{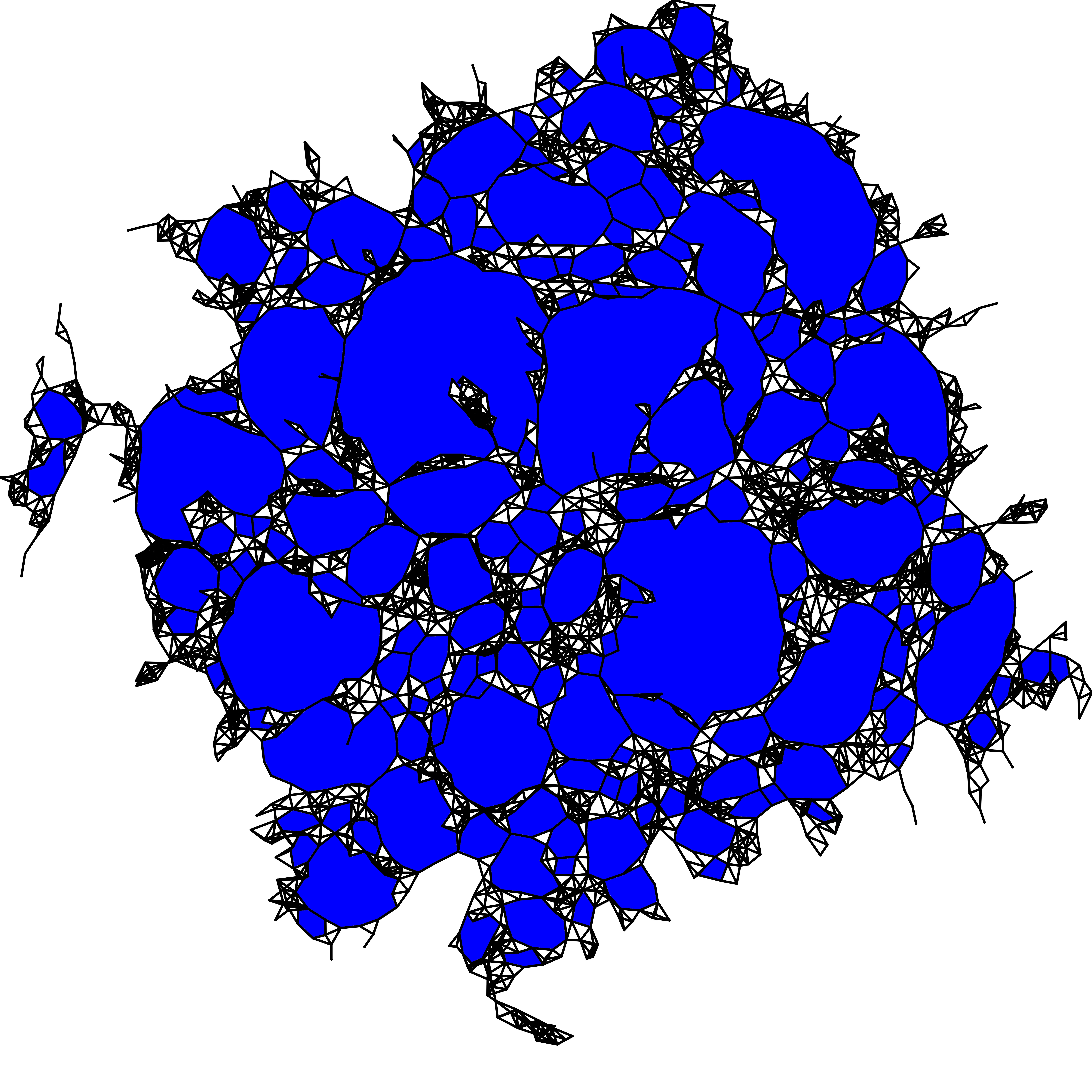}
		\end{minipage}
	}
	\caption{The result of hole detection by FD-CCL and FD-CT.}
	\label{fig:Hole_Detection_Result_Different_Algorithm}
\end{figure}

Figure \ref{fig:Experiment_Workflow} depicts the workflow of the experiments, while Figure \ref{fig:Hole_Detection_Result_Different_Algorithm} shows the results of hole detection using FD-CCL and FD-CT. These methods identify and draw holes whose area exceeds a threshold value calculated based on the canvas size of the FD graph. We found that the performances of FD-CCL and FD-CT are comparable, as both generated the same number of holes, resulting in identical accuracy, sensitivity, and specificity. In Section\ref{subsection:velocity}, we compare FD-CCL with FD-CT by tracing holes and obtaining their properties, such as height, width, area, and coordinates. This experiment was conducted using the OpenCV module with Python. FD-CCL employs the connectedComponentsWithStats method of OpenCV to detect holes and obtain their properties. FD-CT first uses the findContours method to detect holes, then applies the contourArea method to filter the required holes, and finally uses the boundingRect method to obtain the properties of these holes. All experiments were conducted on a computer with an Intel Core i7-6700 CPU @ 3.40 GHz and 16 GB of memory.


\begin{figure}[htp]
	\centering
	\subfigure[Hole detection for sparse graph.]{
		\begin{minipage}[t]{0.45\linewidth}
			\centering
			\includegraphics[width=2in]{Sparse_KKW_n=2000_d=6_Result_CCL.jpg}
		\end{minipage}
	}
	\subfigure[Hole detection for uniform graph.]{
		\begin{minipage}[t]{0.45\linewidth}
			\centering
			\includegraphics[width=2in]{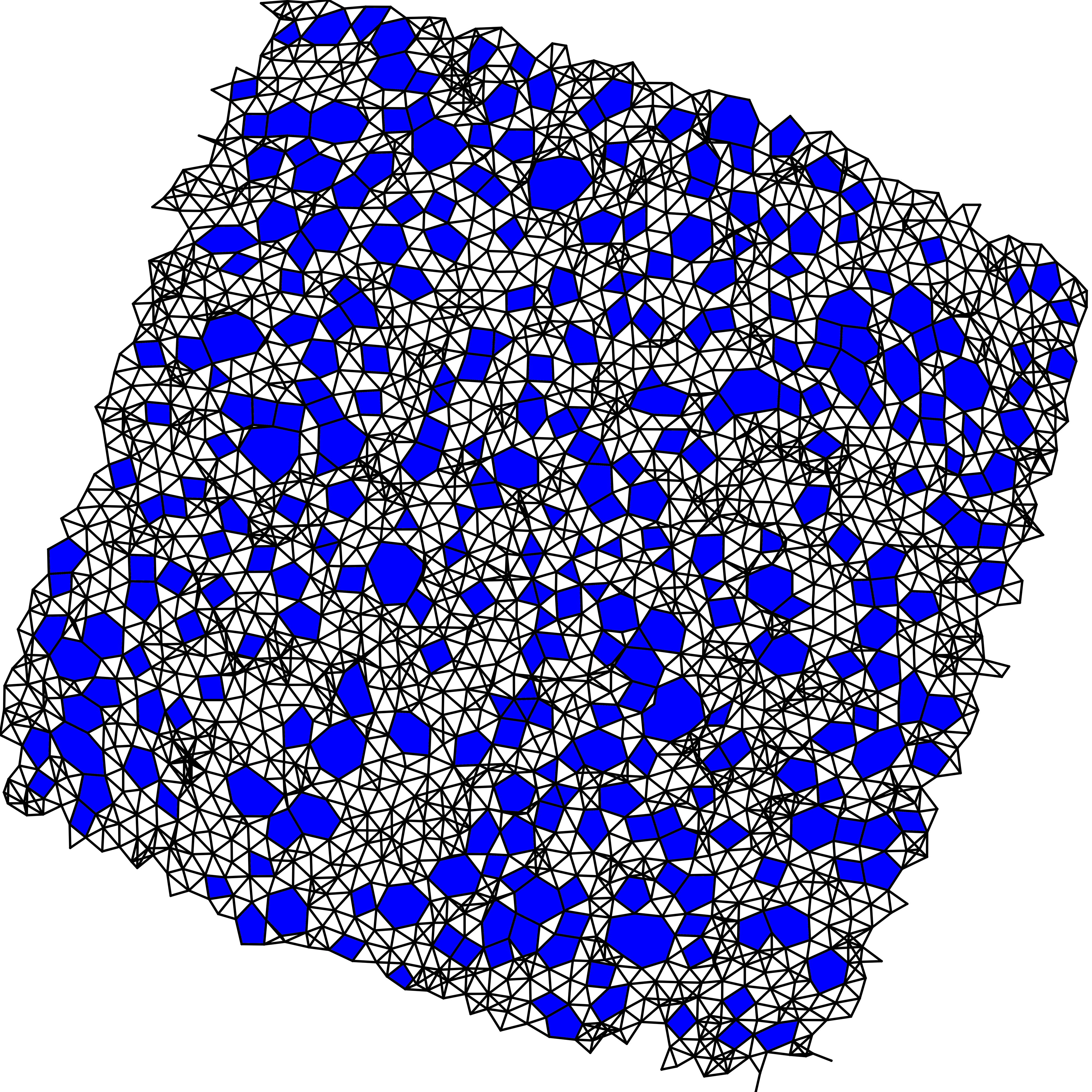}
		\end{minipage}
	}
	\caption{Sample results of hole detection in sparse and uniform layouts.}
	\label{fig:Hole_Detection_Result_Different_Types}
\end{figure}

In the proposed FD-CCL approach, FD algorithms are used to generate the layouts of the Wireless Sensor Networks (WSN) from the connectivity information. Our experiments consider four force-directed graphs algorithms: the Kamada-Kawai (KK) algorithm~\cite{kamada1989algorithm}, the JIGGLE algorithm~\cite{tunkelang1998jiggle}, the ForceAtlas2 (FA2) algorithm~\cite{jacomy2014forceatlas2}, and the Fruchterman Reingold (FR) algorithm~\cite{fruchterman1991graph}. During the execution, the FD algorithm iteratively improves and updates a layout from the given input by adjusting the coordinates of sensors based on force calculations. The program stops updating the layout when it reaches the termination criteria. The generated layout after the final iteration is then used for hole detection by either the CCL or CT algorithms. With a given input topology, the quality of the layout generated by each FD algorithm can vary significantly. Examples of such variations are illustrated in Figure \ref{fig:Hole_Detection_Result_Different_Types}, where different types of force-directed graphs are generated from the same input topology. For the purpose of illustration, the holes from these images are filled with blue color.


To ensure a fair comparison between FD-CCL and FD-CT, we used a default canvas size of 600 $\times$ 600 when generating the layouts. In other words, the sensors are placed on a two-dimensional surface with 600 units in height and width. However, setting a uniform canvas size for the experiment alone may not be sufficient for hole detection since it can result in edge overlapping in the layout, especially when the input topology contains numerous sensors with a high connectivity degree. To address this issue, we use a scaling technique to increase the size of the layout canvas. The functions for scaling are defined as follows:
\begin{equation}
	L_{width}(n)=4 \times n
\end{equation}
\begin{equation}
	L_{height}(n)=4 \times n
\end{equation}
where $L_{width}(n)$ and $L_{height}(n)$ are the solution of the canvas size determination problem. $n$ is the number of sensor nodes in the layout. In the experiments, we used $n$ and $d$ to denote the number of sensors and the connectivity degree of the FD network. The dataset containing nodes' coordinates was generated using four FD algorithms with a ground truth dataset that only contains the connectivity information of nodes. The dataset was pre-processed to remove any outliers or missing values. 

\begin{figure}[H]
	\centering
	\includegraphics[width=0.8\textwidth]{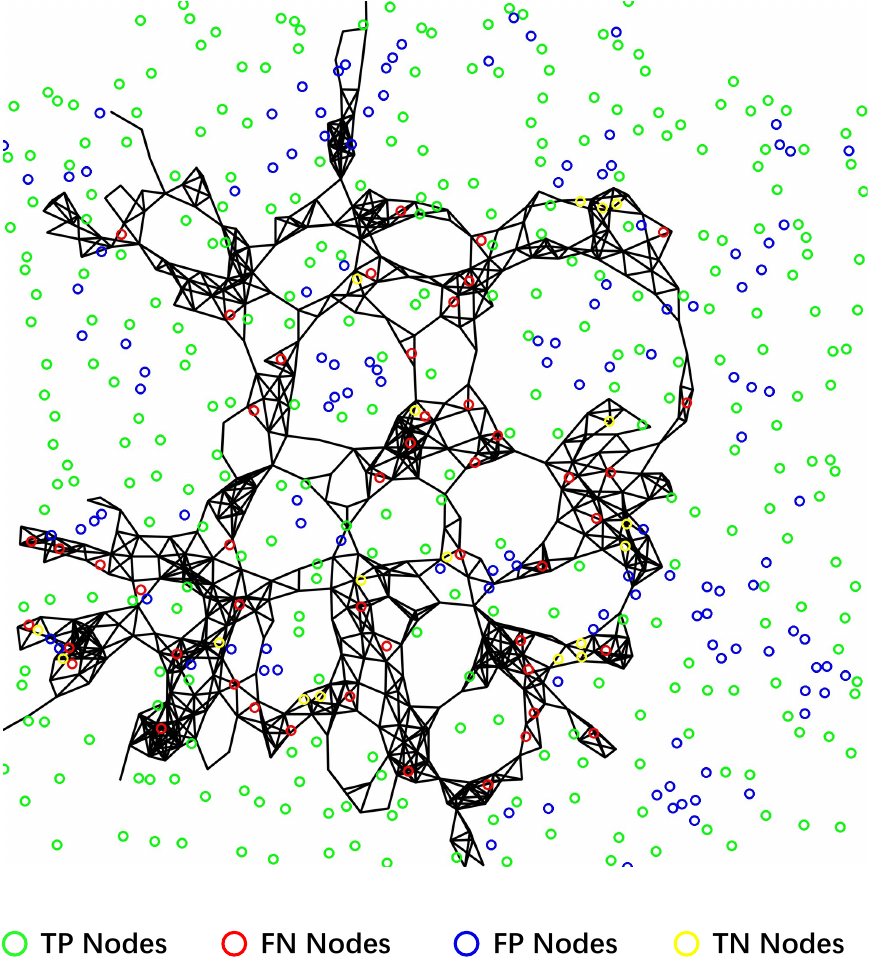}
	\caption{The example of TP, TN, FP, FN in FD-CCL.}
	\label{fig:Example_TP_TN_FP_FN}
\end{figure}

We divide the dataset into five groups by connectivity degrees. When using different FD algorithms with the same values of $n$ and $d$, different layouts may be generated. 
%
In the experiments, we employ sensitivity and specificity criteria for evaluation. Figure \ref{fig:Example_TP_TN_FP_FN} illustrates the examples of True Positive (TP), True Negative (TN), False Positive (FP), and False Negative (FN), and the definitions of TP, TN, FP, and FN are as follows:
\begin{itemize}
	\item [1)]True Positive(TP):
	The number of ground truth nodes on the contours detected by FD-CCL from FD graph.
	\item [2)]True Negative(TN):
	The number of ground truth nodes outside the contours not detected by FD-CCL from FD graph.
	\item [3)]False Positive(FP):
	The number of ground truth nodes outside the contours detected by FD-CCL from FD graph.
	\item [4)]False Negative(FN):
	The number of ground truth nodes on the contours not detected by FD-CCL from FD graph.
\end{itemize}

The formulas for calculating sensitivity and specificity criterion can be defined as follows:
\begin{equation}
	\mbox{Hole Sensitivity}=\frac{TP}{TP+FP}
\end{equation}
\begin{equation}
	\mbox{Hole Specificity}=\frac{TN}{TN+FN}
\end{equation}

\subsection{Feasibility analysis of FD-CCL}\label{subsection:FD-CCL_feasibility}

\begin{figure}[H]
	\centering
	\includegraphics[width=0.6\textwidth]{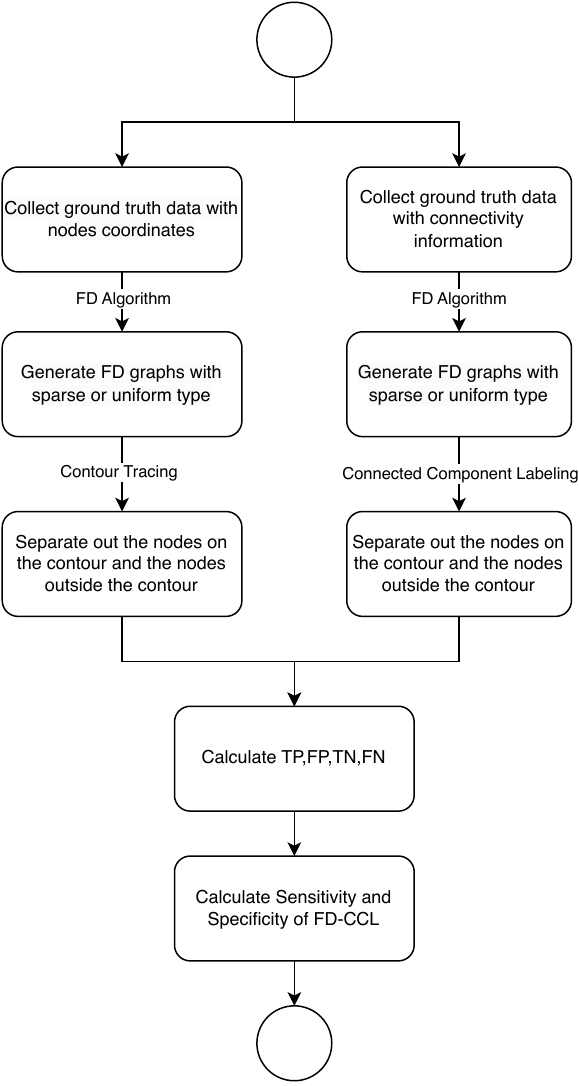}
	\caption{The workflow for calculating sensitivity and specificity of FD-CCL.}
	\label{fig:Feasibility_Workflow}
\end{figure}

This experiment demonstrated the feasibility of FD-CCL by comparing the actual contour tracing result of an FD graph with the result of FD-CCL. Firstly, ground truth data comprising node coordinates was collected in order to generate sparse and uniform FD graphs for contour tracing purposes, with the objective of separating the nodes located on the contour from those situated outside it. Subsequently, ground truth data with connectivity information was gathered in order to generate FD graphs and separate the nodes by FD-CCL. Finally, the TP, FP, TN, FN of FD-CCL were calculated based on the results obtained in the previous two steps. The workflow of this experiment is illustrated in Fig.\ref{fig:Feasibility_Workflow}.

In preparation for the experiment, we used 25 datasets of connectivity information, generated by five different node numbers (200, 500, 1000, 2000, 3000) with five different degrees of connectivity (6, 8, 10, 12, 15) of ground truth data, for the generation of graphs. Four FD algorithms (FR, FA2, KK, JIGGLE) were applied to the aforementioned connectivity information datasets, resulting in the generation of two distinct FD graph layouts (sparse or uniform) for each algorithm.

\subsubsection{Sensitivity and Specificity of FD-CCL for Sparse Layout}\label{subsection:FD-CCL_feasibility_sparse}

We selected the dataset of FD graphs with sparse layout and divided the dataset into five groups according to their connectivity degrees. Each group has 20 sparse FD graphs generated on 5 ground truth datasets with different number of nodes by applying 4 FD algorithm.

\begin{figure}[H]
	\centering
	\subfigure[Degree = 6]{
		\begin{minipage}[t]{0.45\linewidth}
			\centering
			\includegraphics[width=2.0in]{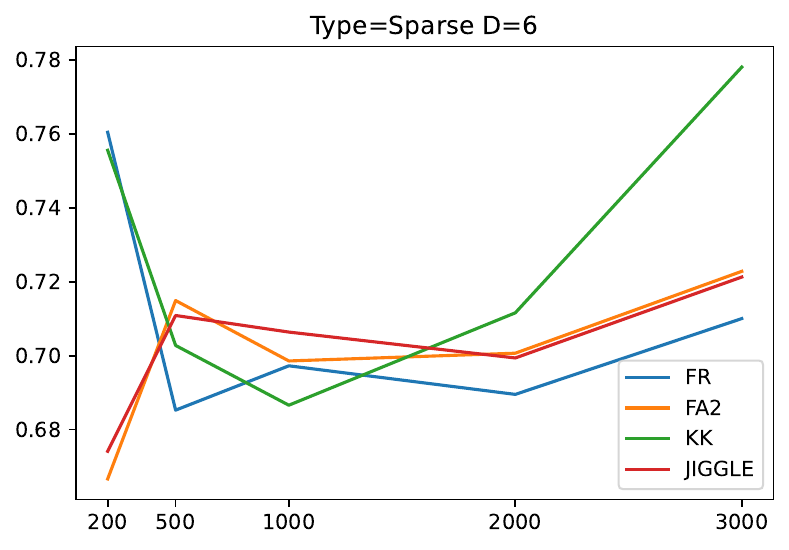}
		\end{minipage}
		\label{fig:Result_Sensitivity_Sparse_d6}
	}
	\subfigure[Degree = 8]{
		\begin{minipage}[t]{0.45\linewidth}
			\centering
			\includegraphics[width=2.0in]{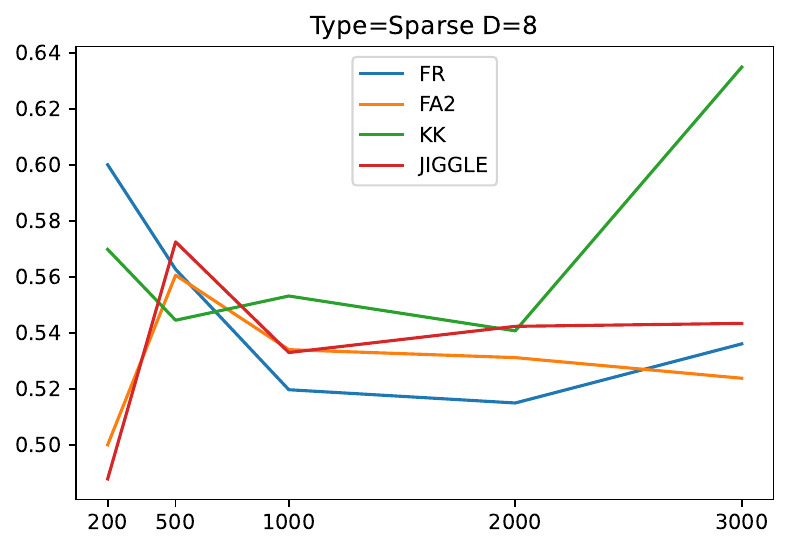}
		\end{minipage}
		\label{fig:Result_Sensitivity_Sparse_d8}
	}
	
	\subfigure[Degree = 10]{
		\begin{minipage}[t]{0.45\linewidth}
			\centering
			\includegraphics[width=2.0in]{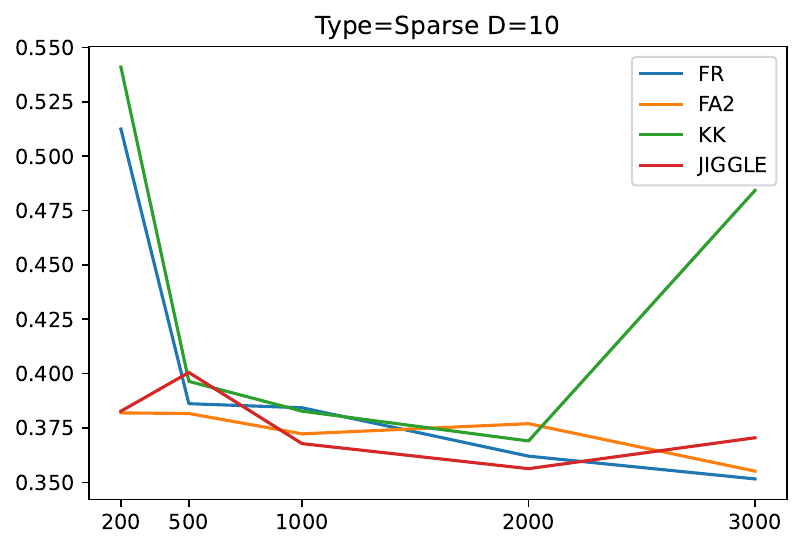}
		\end{minipage}
		\label{fig:Result_Sensitivity_Sparse_d10}
	}
	\subfigure[Degree = 12]{
		\begin{minipage}[t]{0.45\linewidth}
			\centering
			\includegraphics[width=2.0in]{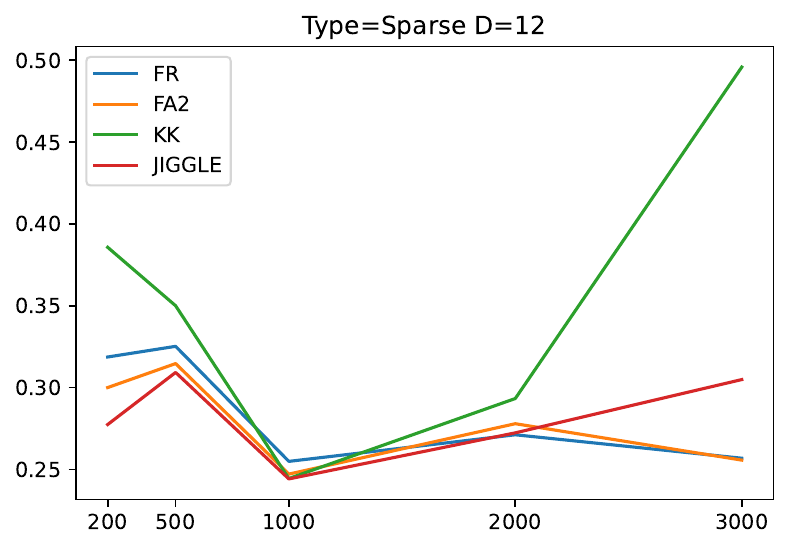}
		\end{minipage}
		\label{fig:Result_Sensitivity_Sparse_d12}
	}
	
	\subfigure[Degree = 15]{
		\begin{minipage}[t]{0.45\linewidth}
			\centering
			\includegraphics[width=2.0in]{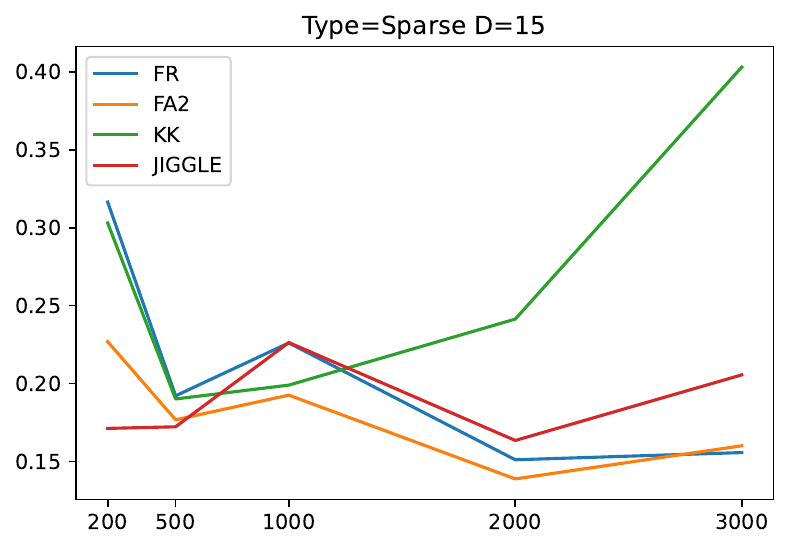}
		\end{minipage}
		\label{fig:Result_Sensitivity_Sparse_d15}
	}
	\centering
	\caption{Sensitivity of FD-CCL in Sparse Layout FD Graph}
	\label{fig:Result_Sensitivity_Sparse}
\end{figure}

\begin{figure}[H]
	\centering
	\subfigure[Degree = 6]{
		\begin{minipage}[t]{0.45\linewidth}
			\centering
			\includegraphics[width=2.0in]{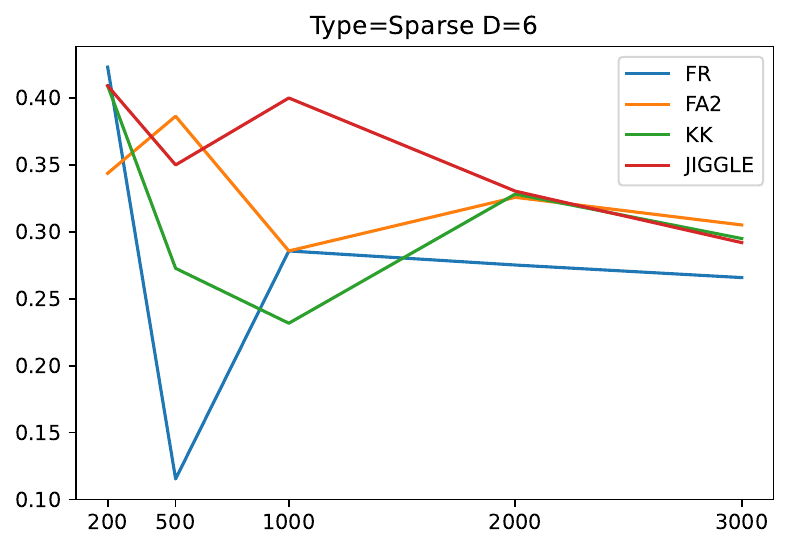}
		\end{minipage}
		\label{fig:Result_Specificity_Sparse_d6}
	}
	\subfigure[Degree = 8]{
		\begin{minipage}[t]{0.45\linewidth}
			\centering
			\includegraphics[width=2.0in]{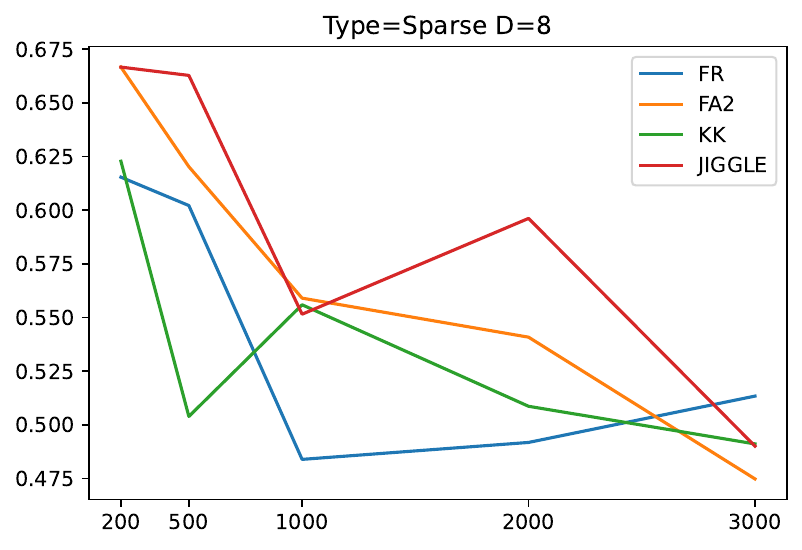}
		\end{minipage}
		\label{fig:Result_Specificity_Sparse_d8}
	}
	
	\subfigure[Degree = 10]{
		\begin{minipage}[t]{0.45\linewidth}
			\centering
			\includegraphics[width=2.0in]{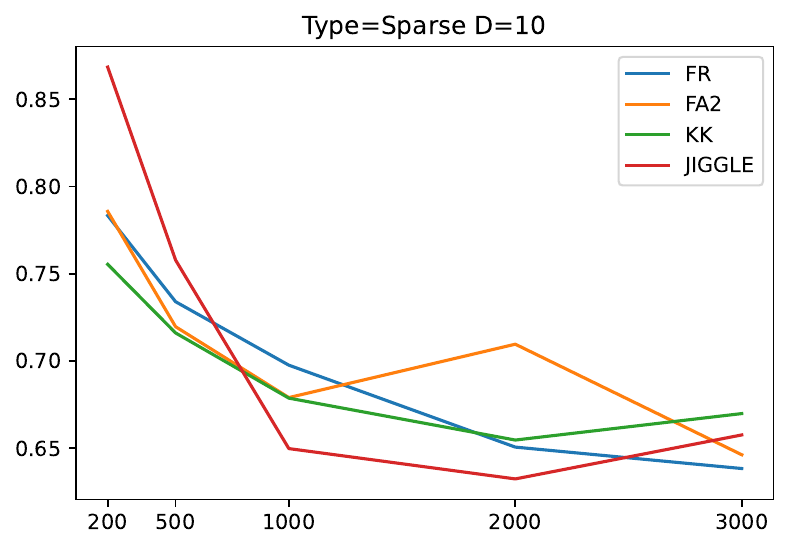}
		\end{minipage}
		\label{fig:Result_Specificity_Sparse_d10}
	}
	\subfigure[Degree = 12]{
		\begin{minipage}[t]{0.45\linewidth}
			\centering
			\includegraphics[width=2.0in]{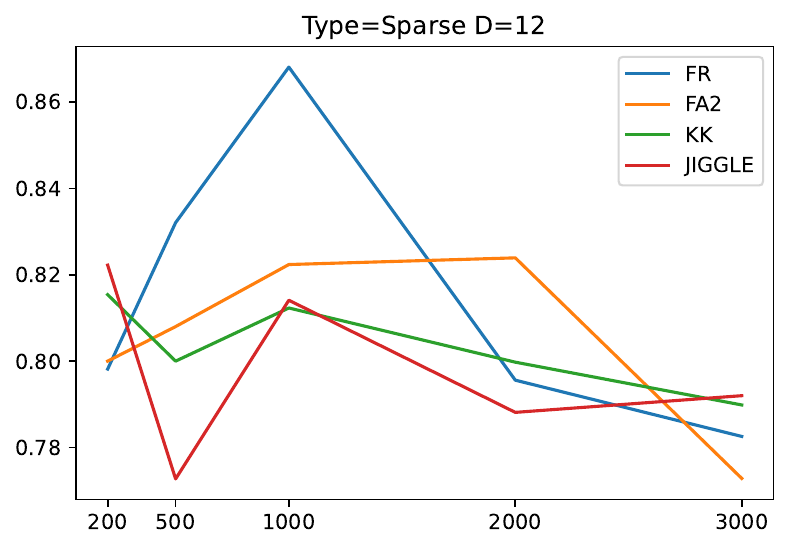}
		\end{minipage}
		\label{fig:Result_Specificity_Sparse_d12}
	}
	
	\subfigure[Degree = 15]{
		\begin{minipage}[t]{0.45\linewidth}
			\centering
			\includegraphics[width=2.0in]{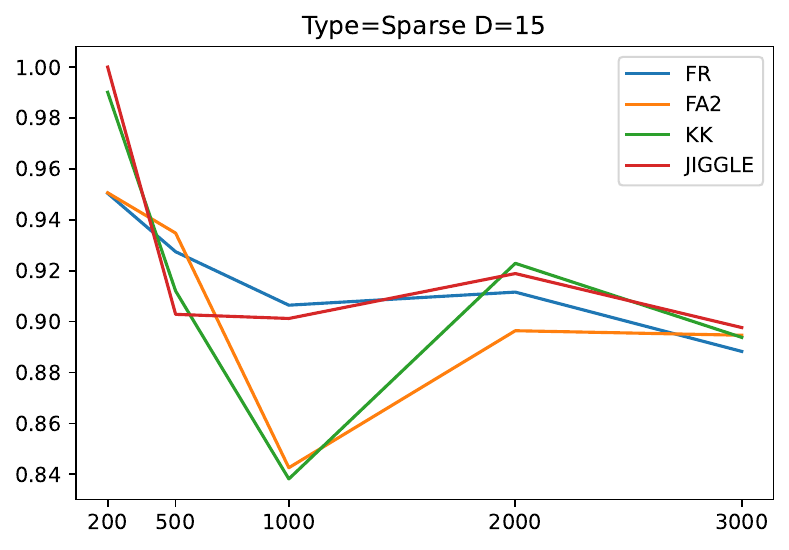}
		\end{minipage}
		\label{fig:Result_Specificity_Sparse_d15}
	}
	\centering
	\caption{Specificity of FD-CCL in Sparse Layout FD Graph}
	\label{fig:Result_Specificity_Sparse}
\end{figure}

To prove the feasibility of FD-CCL, we applied the CCL method to four different types of FD graphs, with each group containing 20 experimental FD graphs. The sensitivity and specificity of FD-CCL for WSN with a sparse layout are shown in Figure \ref{fig:Result_Sensitivity_Sparse} and Figure \ref{fig:Result_Specificity_Sparse}. The results show that the KK Force-Directed algorithm achieves the best performance in hole detection, with a sensitivity about 10\% higher than other FD algorithms. We can also observe from the results that the relationship between the sensitivity of FD-CCL and the degree of connectivity (i.e. graph complexity) is negatively correlated, while the relationship between the specificity of FD-CCL and the degree of connectivity is positively correlated.

\subsubsection{Sensitivity and Specificity of FD-CCL for Uniform Layout}\label{subsection:FD-CCL_feasibility_uniform}
\begin{figure}[H]
	\centering
	\subfigure[Degree = 6]{
		\begin{minipage}[t]{0.45\linewidth}
			\centering
			\includegraphics[width=2.0in]{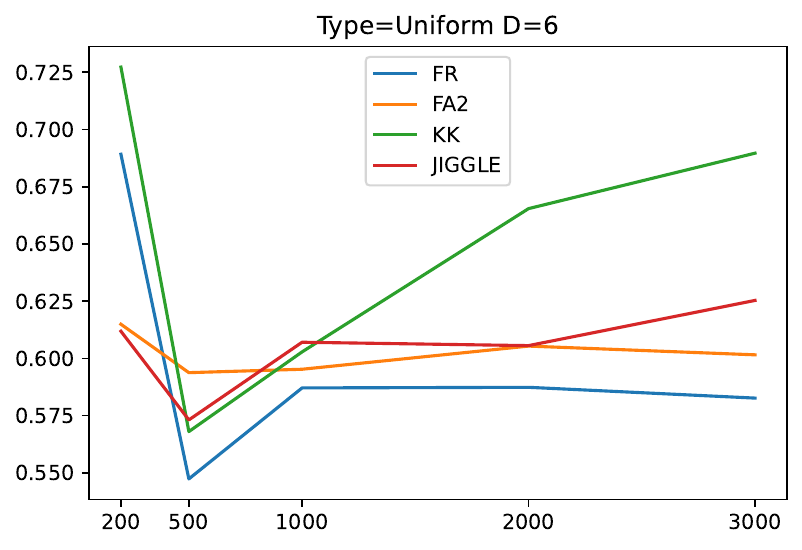}
		\end{minipage}
		\label{fig:Result_Sensitivity_Uniform_d6}
	}
	\subfigure[Degree = 8]{
		\begin{minipage}[t]{0.45\linewidth}
			\centering
			\includegraphics[width=2.0in]{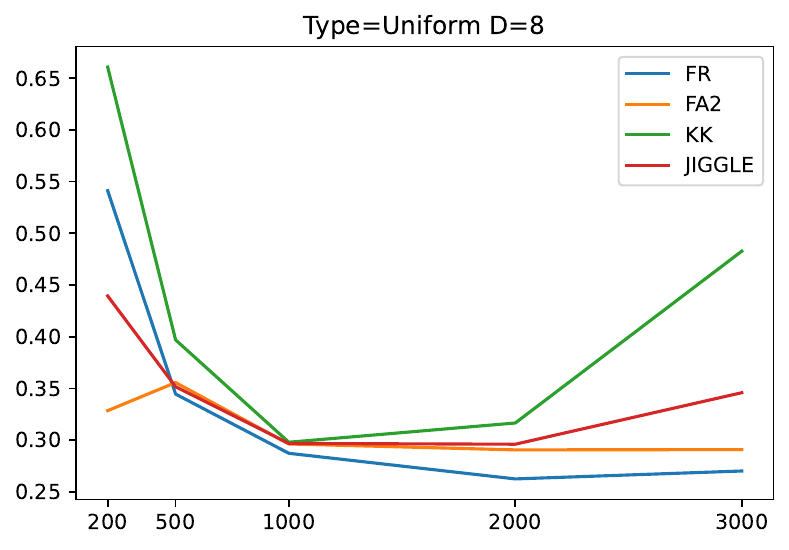}
		\end{minipage}
		\label{fig:Result_Sensitivity_Uniform_d8}
	}
	
	\subfigure[Degree = 10]{
		\begin{minipage}[t]{0.45\linewidth}
			\centering
			\includegraphics[width=2.0in]{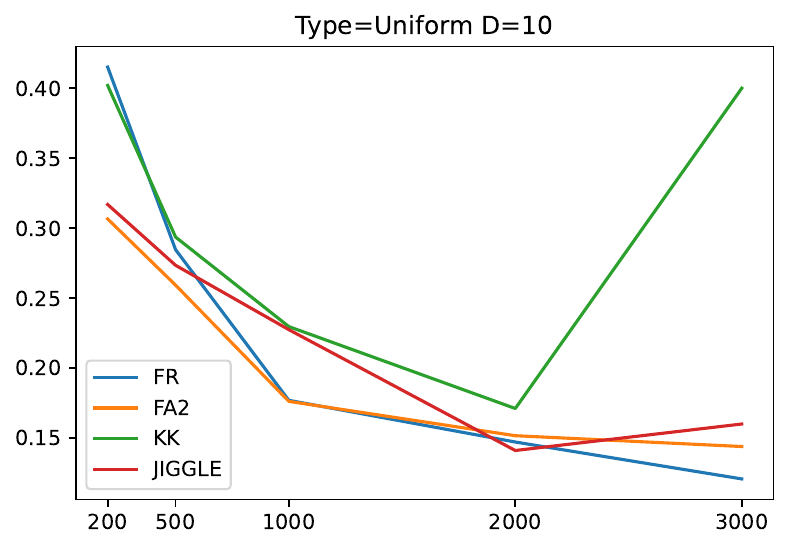}
		\end{minipage}
		\label{fig:Result_Sensitivity_Uniform_d10}
	}
	\subfigure[Degree = 12]{
		\begin{minipage}[t]{0.45\linewidth}
			\centering
			\includegraphics[width=2.0in]{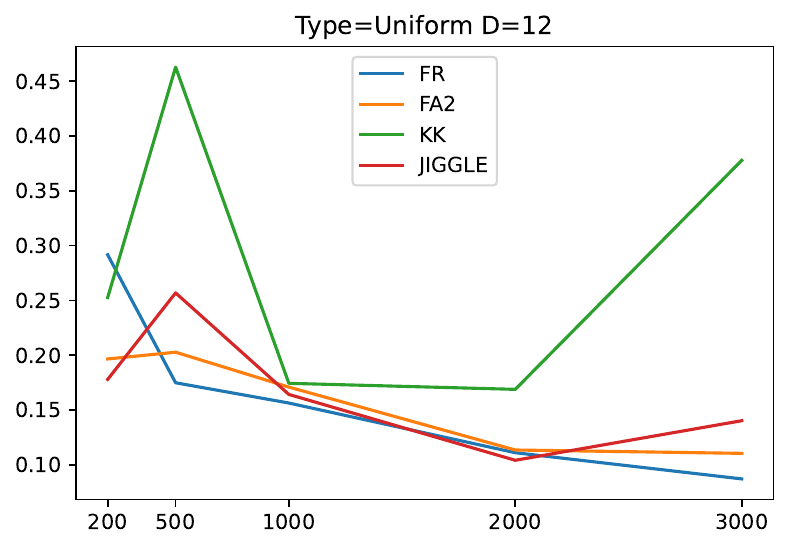}
		\end{minipage}
		\label{fig:Result_Sensitivity_Uniform_d12}
	}
	
	\subfigure[Degree = 15]{
		\begin{minipage}[t]{0.45\linewidth}
			\centering
			\includegraphics[width=2.0in]{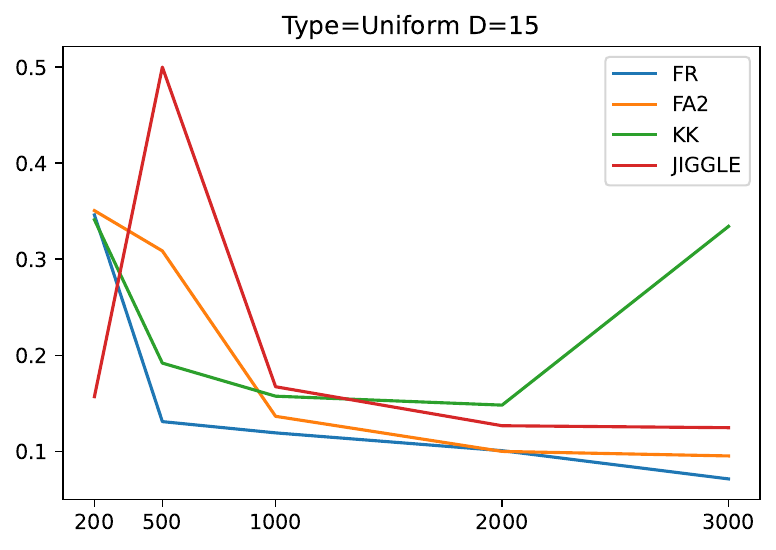}
		\end{minipage}
		\label{fig:Result_Sensitivity_Uniform_d15}
	}
	\centering
	\caption{Sensitivity of FD-CCL in Uniform Layout FD Graph}
	\label{fig:Result_Sensitivity_Uniform}
\end{figure}

\begin{figure}[H]
	\centering
	\subfigure[Degree = 6]{
		\begin{minipage}[t]{0.45\linewidth}
			\centering
			\includegraphics[width=2.0in]{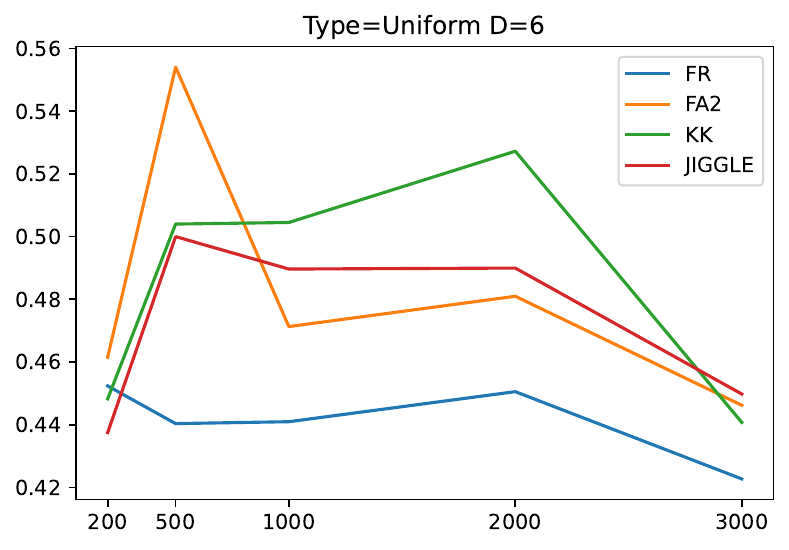}
		\end{minipage}
		\label{fig:Result_Specificity_Uniform_d6}
	}
	\subfigure[Degree = 8]{
		\begin{minipage}[t]{0.45\linewidth}
			\centering
			\includegraphics[width=2.0in]{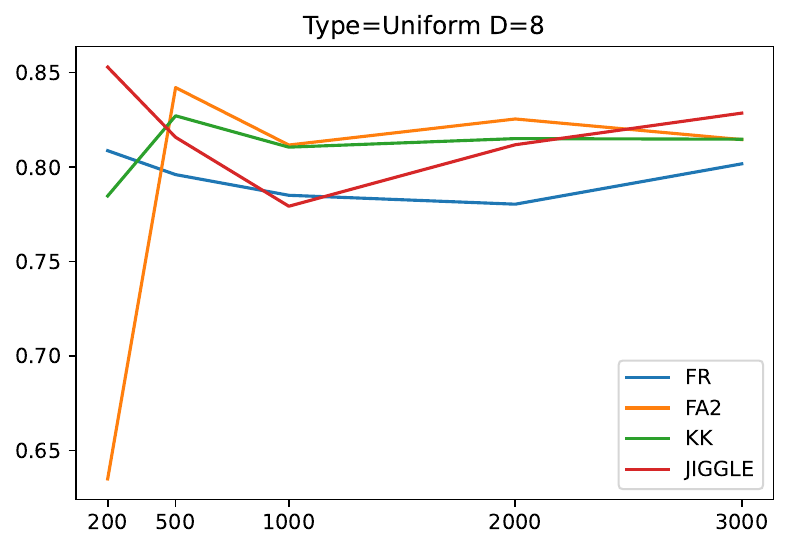}
		\end{minipage}
		\label{fig:Result_Specificity_Uniform_d8}
	}
	
	\subfigure[Degree = 10]{
		\begin{minipage}[t]{0.45\linewidth}
			\centering
			\includegraphics[width=2.0in]{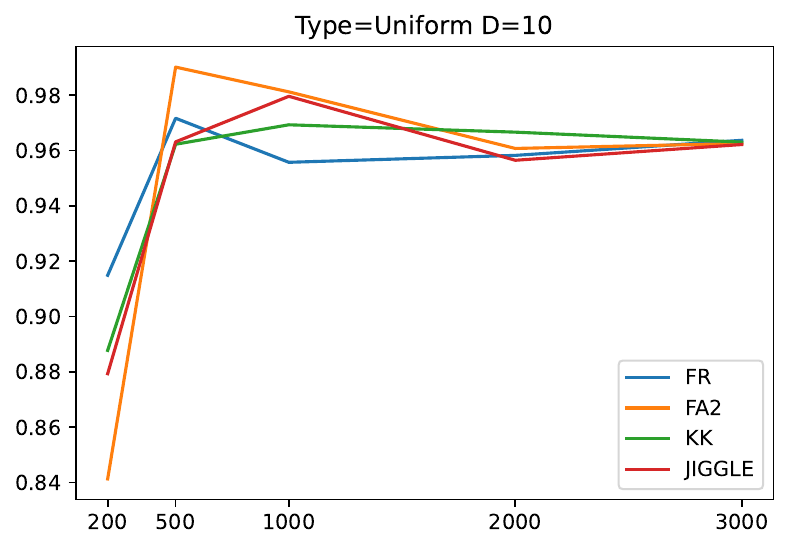}
		\end{minipage}
		\label{fig:Result_Specificity_Uniform_d10}
	}
	\subfigure[Degree = 12]{
		\begin{minipage}[t]{0.45\linewidth}
			\centering
			\includegraphics[width=2.0in]{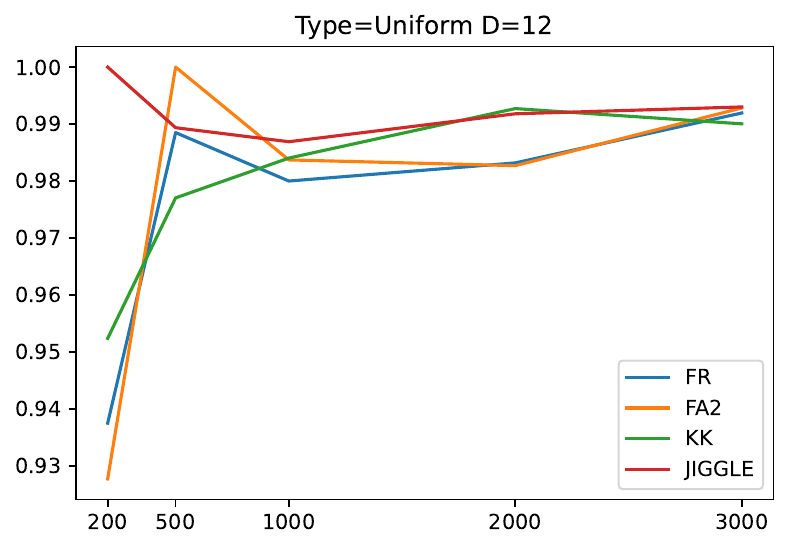}
		\end{minipage}
		\label{fig:Result_Specificity_Uniform_d12}
	}
	
	\subfigure[Degree = 15]{
		\begin{minipage}[t]{0.45\linewidth}
			\centering
			\includegraphics[width=2.0in]{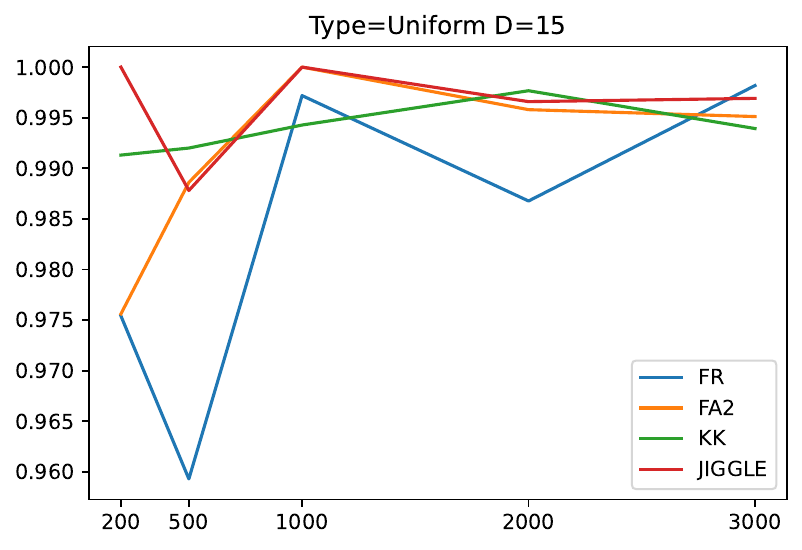}
		\end{minipage}
		\label{fig:Result_Specificity_Uniform_d15}
	}
	\centering
	\caption{Specificity of FD-CCL in Uniform Layout FD Graph}
	\label{fig:Result_Specificity_Uniform}
\end{figure}

The sensitivity and specificity of FD-CCL for WSNs with uniform layouts are shown in Figure \ref{fig:Result_Sensitivity_Uniform} and Figure \ref{fig:Result_Specificity_Uniform}. Compared to the sensitivity and specificity of FD-CCL for WSNs with a sparse layout, FD-CCL performs better when processing a uniform layout, with about 10 percent higher sensitivity than when processing a sparse layout.

\subsection{Comparison of FD-CCL and FD-CT for Locating Holes}\label{subsection:velocity}
\begin{figure}[H]
	\centering
	\subfigure[Degree = 6]{
		\begin{minipage}[t]{0.45\linewidth}
			\centering
			\includegraphics[width=2.0in]{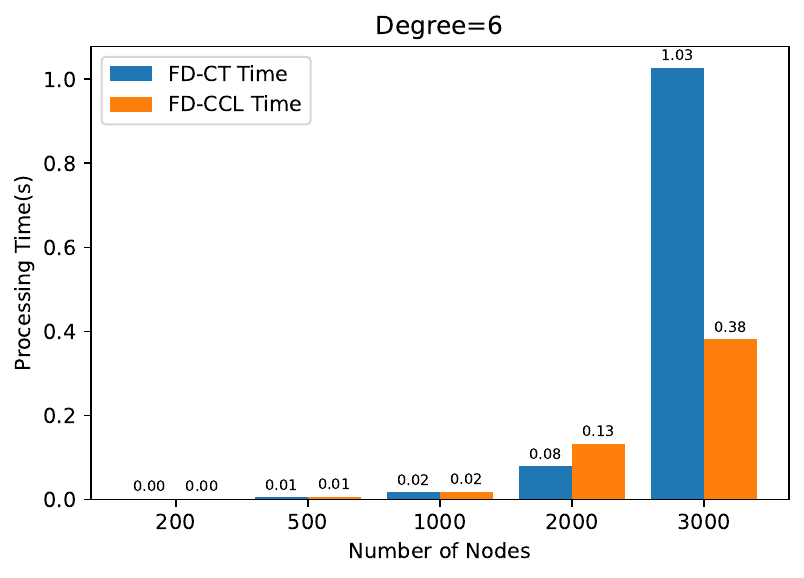}
		\end{minipage}
		\label{fig:Result_LocatingVelocity_d6}
	}
	\subfigure[Degree = 8]{
		\begin{minipage}[t]{0.45\linewidth}
			\centering
			\includegraphics[width=2.0in]{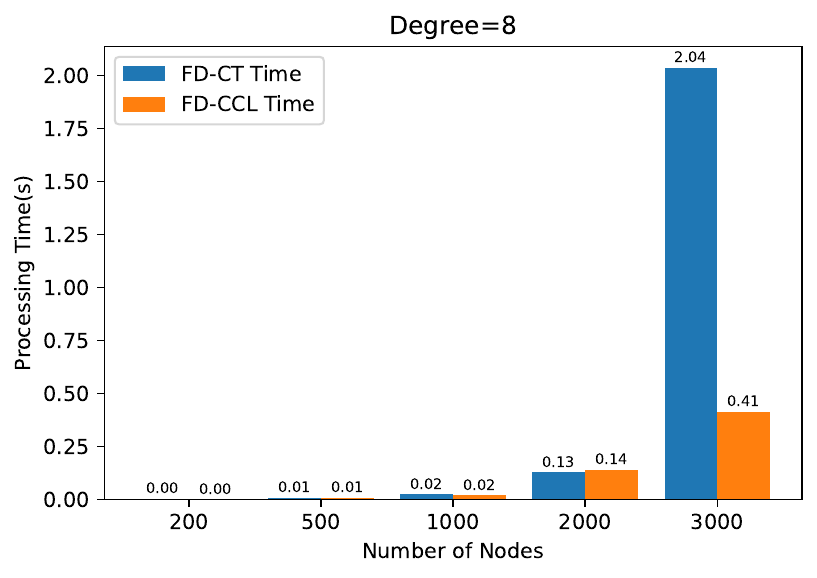}
		\end{minipage}
		\label{fig:Result_LocatingVelocity_d8}
	}
	
	\subfigure[Degree = 10]{
		\begin{minipage}[t]{0.45\linewidth}
			\centering
			\includegraphics[width=2.0in]{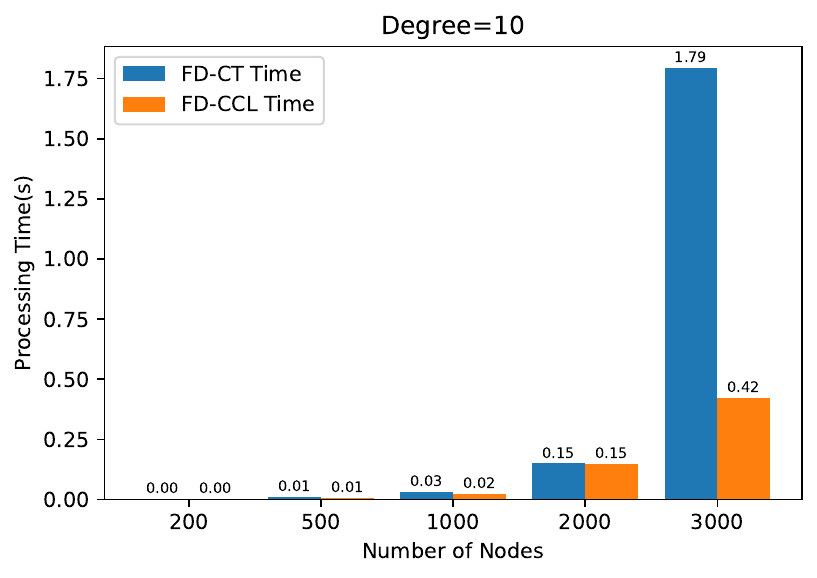}
		\end{minipage}
		\label{fig:Result_LocatingVelocity_d10}
	}
	\subfigure[Degree = 12]{
		\begin{minipage}[t]{0.45\linewidth}
			\centering
			\includegraphics[width=2.0in]{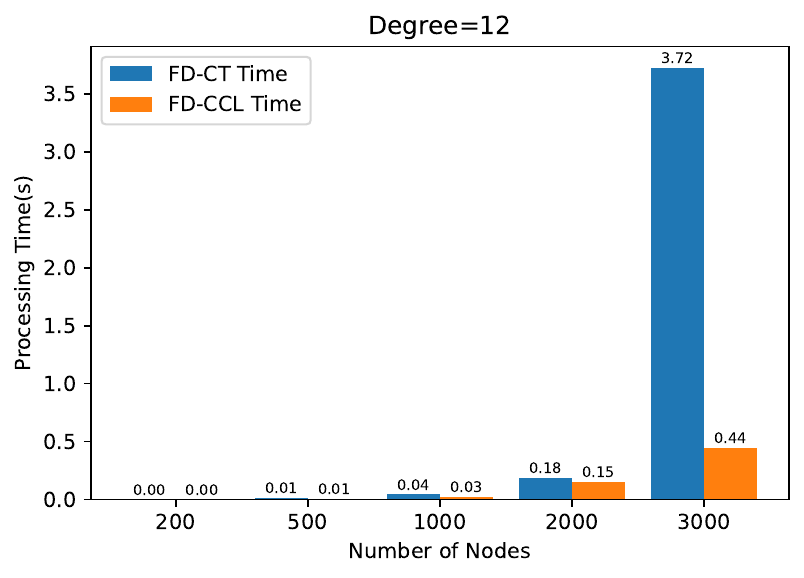}
		\end{minipage}
		\label{fig:Result_LocatingVelocity_d12}
	}
	
	\subfigure[Degree = 15]{
		\begin{minipage}[t]{0.45\linewidth}
			\centering
			\includegraphics[width=2.0in]{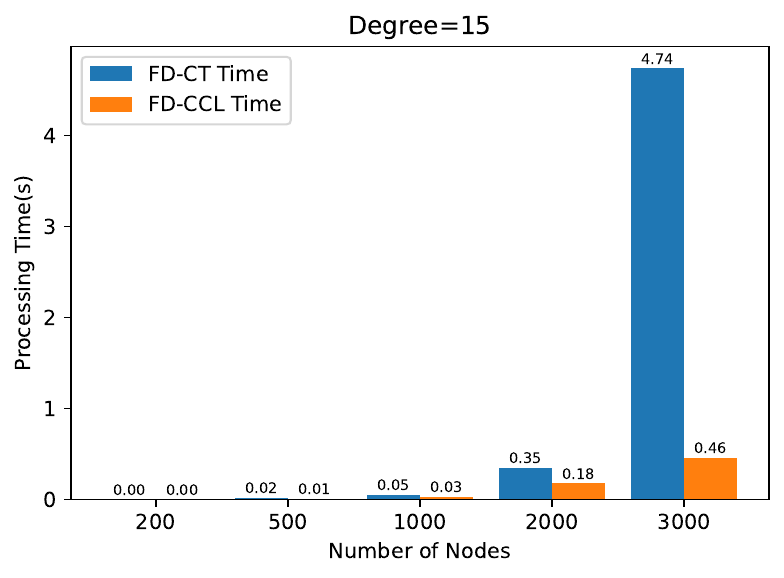}
		\end{minipage}
		\label{fig:Result_LocatingVelocity_d15}
	}
	\centering
	\caption{Comparison of efficiency between FD-CCL and FD-CT}
	\label{fig:Result_LocatingVelocity}
\end{figure}

In this experiment, we used the processing time, a common metric in benchmarking practice, to evaluate the effectiveness of FD-CCL and FD-CT for locating holes. To have a fair comparison, for both FD-CCL and FD-CT, the FD layouts tested in the experiment have same the average degree and the number of nodes. First, we used the CT algorithm and CCL algorithm to process the FD graph dataset, which is divided into five groups by connectivity degrees, with each group containing 40 different graphs for the experiment. Figure \ref{fig:Result_LocatingVelocity} shows the result of processing time comparison between FD-CCL and FD-CT for locating holes. When processing complex graphs, which consist of more nodes and higher connectivity degrees, FD-CCL performs better. When $d$=6, FD-CCL is about 3 times faster than FD-CT, especially at $d$=15, FD-CCL is about 10 times faster than FD-CT.

The reasons why FD-CCL is more efficient in detecting and locating holes with their properties can be summarized as follow. According to Algorithm \ref{alg:CCL}, 
FD-CCL calculates and stores hole information using component labels during the hole detection process. FD-CCL labels the entire region of holes, whereas FD-CT traces the boundary of holes and only records the pixels of the hole outlines. Therefore, FD-CT requires additional calculations. 
Furthermore, 
during the hole filtering process, FD-CT needs to calculate the area of the hole based on a  threshold. In contrast, FD-CCL directly obtains the hole area from the processing results.

\section{Conclusion}
\label{section6}
Hole detection is a crucial task in Wireless Sensor Networks (WSN) as it enables reliable data collection and communication. In a WSN, nodes are deployed in a specific area to monitor and collect data. However, due to various reasons such as battery or hardware failures, some nodes may become isolated, creating gaps or holes in the network. These holes can significantly impact the network's performance by limiting data transmission, reducing coverage, and compromising the quality of collected data. Therefore, in this paper, we propose a novel approach for detecting and locating holes in Wireless Sensor Networks (WSN) called Connected Component Labeling with Force-Directed algorithm (FD-CCL). FD-CCL is a coordinates-free algorithm that relies on the Force-Directed (FD) algorithm and does not require any additional physical information. It comprises two main phases: WSN layout generation and hole detection. The FD algorithm generates a layout from the connectivity information, and the CCL contour tracing method detects holes in the FD graph. The FD algorithm is a graph visualization algorithm that creates visually appealing representations of graphs, while the CCL contour tracing method is a pixel-following algorithm in the field of Computer Vision with a time complexity of $O(N)$, where $N$ is the number of nodes.

In this paper, the hole sensitivity and specificity of FD-CCL are presented, proving the feasibility of the FD-CCL algorithm. Moreover, through comparisons with the FD-based state-of-the-art hole detection approach , we show that FD-CCL is more efficient in terms of processing time for locating and obtaining the properties of holes. Given the same average degree and number of nodes in WSNs, FD-CCL performs 3 times faster than FD-CT for locating and obtaining the properties of holes.

\section*{Acknowledgments}
This research was funded by the University of Macau (file no. MYRG2022-00162-FST and MYRG2019-00136-FST).

\section*{CRediT author statement}
Jiacheng Xu: Conceptualization, Methodology, Software, Writing- Original draft preparation.  
Hou-Wan Long: Writing- Reviewing and Editing.
Cheong Se-Hang: Methodology, Software.
Yain-Whar Si: Supervision, Writing- Reviewing and Editing, Funding acquisition.


\end{document}